\documentclass{tlp}
\usepackage{aopmath}

\usepackage{epsfig}
\usepackage{lscape}

\usepackage[english]{babel}

\usepackage{amsfonts}
\usepackage{amssymb}

\renewcommand{\ttdefault}{cmtt}

\newtheorem{definition}{Definition} 
\newtheorem{example}{Example} 

\let\from=\leftarrow

\begin{document}
\bibliographystyle{acmtrans}

\long\def\comment#1{}

\title[ILP in Databases: from \textsc{Datalog} to \textsc{$\mathcal{DL}$+log}$^{\neg\vee}$]{Inductive Logic Programming in Databases:\\from \textsc{Datalog} to \textsc{$\mathcal{DL}$+log}$^{\neg\vee}$}

\author[F.A. Lisi]
{FRANCESCA A. LISI \\
Dipartimento di Informatica \\
Universit\`{a} degli Studi di Bari \\
Via Orabona 4\\
70125 Bari, Italy\\
E-mail: lisi@di.uniba.it
}

\pagerange{\pageref{firstpage}--\pageref{lastpage}}
\volume{\textbf{10} (3):}
\jdate{September 2007}
\setcounter{page}{1}
\pubyear{2010}
\submitted{24 April 2009}
\revised {15 November 2009}
\accepted{1 January 2010}

\maketitle

\label{firstpage}

\begin{abstract}
In this paper we address an issue that has been brought to the attention of the database community with the advent of the Semantic Web, i.e. the issue of how ontologies (and semantics conveyed by them) can help solving typical database problems, through a better understanding of KR aspects related to databases. In particular, we investigate this issue from the ILP perspective by considering two database problems, (i) the definition of views and (ii) the definition of constraints, for a database whose schema is represented also by means of an ontology. Both can be reformulated as ILP problems and can benefit from the expressive and deductive power of the KR framework \textsc{$\mathcal{DL}$+log}$^{\neg\vee}$. We illustrate the application scenarios by means of examples.
\end{abstract}


\section{Motivation}\label{sect:motivation}

Inductive Logic Programming (ILP) has been historically concerned with the induction of rules from examples for classification purposes \cite{Nienhuys97}. Due to the close relation between Logic Programming and Relational Databases \cite{Ceri90}, ILP has established itself as a major approach to Relational Data Mining \cite{DzeroskiL01}. Indeed, \textsc{Datalog} \cite{CeriGT89} is the most widely used Knowledge Representation (KR) framework in ILP. Conversely, interesting extensions of \textsc{Datalog} such as $\textsc{Datalog}^{\neg \vee}$ \cite{EiterGM97} have attracted very little attention in ILP. Some effort has been made also at making ILP more able to face the challenges posed by Relational Data Mining applications, e.g. scalability \cite{Blockeel99}. However the actual added value of ILP with respect to far more efficient approaches still remains the use of prior conceptual knowledge (also known as \emph{background knowledge}, or shortly BK) during the learning process which enables the induction of conceptually meaningful rules. Yet, the BK in ILP is often not organized around a well-formed conceptual model. This practice seems to ignore the latest achievements in conceptual modeling such as ontologies. 

In Artificial Intelligence, an \emph{ontology} refers to an engineering artifact (more precisely, produced according to the principles of \emph{Ontological Engineering} \cite{GomezPerez04}), constituted by a specific vocabulary used to describe a certain reality, plus a set of explicit assumptions regarding the
intended meaning of the vocabulary words. This set of assumptions has usually the form of a first-order logical (FOL) theory, where vocabulary words appear as unary or binary
predicate names, respectively called concepts and relations. 
More formally, an ontology
is a formal explicit specification of a shared conceptualization for a domain of
interest \cite{Gruber93}. Among the other things, this definition emphasizes the fact that an ontology
has to be specified in a language that comes with a formal semantics. Only by
using such a formal approach ontologies provide the machine interpretable meaning
of concepts and relations that is expected when using an ontology-based approach. Among the formalisms proposed by Ontological Engineering, the most currently used are \emph{Description Logics} (DLs) \cite{BaaderCMcGNPS07}. In particular, the advent of the Semantic Web \cite{Berners-Lee01} has given a tremendous impulse to research on DL-based ontology languages. Indeed the DL $\mathcal{SHIQ}$ \cite{HorrocksST00} has been the starting point for the definition of the W3C standard mark-up language OWL \cite{HorrocksP-SvH03}. Note that DLs are decidable fragments of FOL that are incomparable with Clausal Logics (CLs) as regards the expressive power \cite{Borgida96} and the semantics \cite{Rosati05-ppswr}. Yet, DLs and CLs can be combined according to some limited forms of hybridization. E.g., \textsc{$\mathcal{DL}$+log}$^{\neg\vee}$ is a general KR framework that allows for the tight integration of DLs and $\textsc{Datalog}^{\neg \vee}$ by imposing the condition of weak $\mathcal{DL}$-safeness on hybrid rules \cite{Rosati06}\footnote{We prefer to use the name \textsc{$\mathcal{DL}$+log}$^{\neg\vee}$ instead of the original one $\mathcal{DL}$+\textsc{log} in order to emphasize the $\textsc{Datalog}^{\neg\vee}$ component of the framework.}. We argue that the adoption of such hybrid KR systems can help overcoming the current difficulties in accommodating ontologies in ILP. 

In this paper we address an issue that has been brought to the attention of the database community with the advent of the Semantic Web, i.e. the issue of how ontologies (and semantics conveyed by them) can help solving typical database problems, through a better understanding of KR aspects related to databases. In particular, we investigate this issue from the ILP perspective by considering two database problems: 
\begin{itemize}
	\item the definition of views
	\item the definition of constraints
\end{itemize}
for a database whose schema is represented also by means of an ontology. Both can be reformulated as ILP problems and can benefit from the expressive and deductive power of the KR framework \textsc{$\mathcal{DL}$+log}$^{\neg\vee}$, mainly from its nonmonotonic (NM) features. We illustrate the application scenarios by means of examples.
 
The paper is organized as follows. Section \ref{sect:background} provides basic notions on DLs, a short summary of KR research on the integration of DLs and CLs, and a brief introduction to ILP. Section \ref{sect:dl+log} introduces syntax, semantics and reasoning of \textsc{$\mathcal{DL}$+log$^{\neg\vee}$}. Section \ref{sect:inducing-views} and Section \ref{sect:inducing-constraints} define the ILP proposals for inducing database views and database constraints, respectively, within the \textsc{$\mathcal{DL}$+log$^{\neg\vee}$} framework. Section \ref{sect:rel-work} surveys related work. Section \ref{sect:concl} concludes the paper with final remarks.

\section{Background}\label{sect:background}

\subsection{Representing ontologies}\label{sect:dl}

\begin{table*}[t]
\small
\caption{Syntax and semantics of some typical DL constructs.}\label{tab:DL}
\begin{center}
\begin{tabular}{r@{\hspace{.15cm}}c@{\hspace{.15cm}}l}
\hline
\noalign{\smallskip}
bottom (resp. top) concept & $\bot$ (resp. $\top$) & $\emptyset$ (resp. $\Delta^\mathcal{I}$)\\
atomic concept & $A$ & $A^\mathcal{I} \subseteq \Delta^\mathcal{I}$\\
(abstract) simple role & $S$ & $S^\mathcal{I} \subseteq \Delta^\mathcal{I} \times \Delta^\mathcal{I}$\\
(abstract) individual & $a$ & $a^\mathcal{I} \in \Delta^\mathcal{I}$\\
\hline
\noalign{\smallskip}
concept & $C$\\
role & $R$\\
concept negation & $\neg C$ & $\Delta^\mathcal{I} \setminus C^\mathcal{I}$  \\
concept intersection & $C_1 \sqcap C_2$ & $C^\mathcal{I}_1 \cap C^\mathcal{I}_2$\\
concept union & $C_1 \sqcup C_2$ & $C^\mathcal{I}_1 \cup C^\mathcal{I}_2$\\
value restriction & $\forall R.C$ & $\{x \in \Delta^\mathcal{I} \mid \forall y\ (x,y)\in R^\mathcal{I} \rightarrow y \in
C^\mathcal{I}\}$\\
existential restriction & $\exists R.C$ & $\{x \in \Delta^\mathcal{I} \mid \exists y\ (x,y)\in R^\mathcal{I} \wedge y \in
C^\mathcal{I}\}$\\
at least number restriction & $\geq n R$ & $\{x \in \Delta^\mathcal{I} \mid | \{y| (x,y)\in R^\mathcal{I}\} | \geq n\}$\\
at most number restriction & $\leq n R$ & $\{x \in \Delta^\mathcal{I} \mid | \{y| (x,y)\in R^\mathcal{I}\} | \leq n\}$\\
at least qualif. number restriction & $\geq n R.C$ & $\{x \in \Delta^\mathcal{I} \mid | \{y \in
C^\mathcal{I} | (x,y)\in R^\mathcal{I}\} | \geq n\}$\\
at most qualif. number restriction & $\leq n R.C$ & $\{x \in \Delta^\mathcal{I} \mid | \{y \in
C^\mathcal{I} | (x,y)\in R^\mathcal{I}\} | \leq n\}$\\
role inversion & $R^{-}$ & $\{(x,y) \in \Delta^\mathcal{I} \times \Delta^\mathcal{I} \mid  (y,x)\in R^\mathcal{I}\}$\\
role intersection & $R_1 \sqcap R_2$ & $R^\mathcal{I}_1 \cap R^\mathcal{I}_2$\\
\hline
\end{tabular}
\end{center}
\end{table*}

DLs are a family of decidable FOL fragments that allow for the specification of knowledge in terms of classes (\emph{concepts}), 
instances (\emph{individuals}), and binary relations between instances (\emph{roles}) \cite{Borgida96}. Complex concepts can be defined from atomic concepts and roles by means of constructors. Syntax and semantics of some typical DL constructs are reported in Table \ref{tab:DL}. E.g., concept descriptions in the basic DL $\mathcal{AL}$ are formed according to only the constructors of atomic negation, concept conjunction, value restriction, and limited existential restriction. The DLs $\mathcal{ALC}$ and $\mathcal{ALN}$ are members of the $\mathcal{AL}$ family. The former extends $\mathcal{AL}$ with (arbitrary) concept negation (also called complement and equivalent to having both concept union and full existential restriction), whereas the latter with number restriction. The DL $\mathcal{ALCNR}$ adds to the constructors inherited from $\mathcal{ALC}$ and $\mathcal{ALN}$ a further one: role intersection. Conversely, in the DL $\mathcal{SHIQ}$ \cite{HorrocksST00} it is allowed to invert roles and to express qualified number restrictions of the form $\geq n R.C$ and $\leq n R.C$ where $R$ is a simple role. Also transitivity holds for roles. A role (expression) is called complex if it contains any role operations other than inversion, e.g. role intersection.

\begin{table*}[t]
\small
\caption{Syntax and semantics of DL KBs.}\label{tab:DL-KB}
\begin{center}
\begin{tabular}{r@{\hspace{.15cm}}c@{\hspace{.15cm}}l}
\hline
\noalign{\smallskip}
concept equivalence axiom & $C_1 \equiv C_2$ & $C^\mathcal{I}_1 = C^\mathcal{I}_2$  \\
concept subsumption axiom & $C_1 \sqsubseteq C_2$ & $C^\mathcal{I}_1 \subseteq C^\mathcal{I}_2$\\
\hline
\noalign{\smallskip}
role equivalence axiom & $R_1 \equiv R_2$ & $R^\mathcal{I}_1 = R^\mathcal{I}_2$  \\
role inclusion axiom & $R_1 \sqsubseteq R_2$ & $R^\mathcal{I}_1 \subseteq R^\mathcal{I}_2$\\
\hline
\noalign{\smallskip}
concept assertion & $C(a)$ & $a^\mathcal{I} \in C^\mathcal{I}$\\
role assertion & $R(a,b)$ & $(a^\mathcal{I}, b^\mathcal{I}) \in R^\mathcal{I}$\\
individual equality assertion & $a \approx b$ & $a^\mathcal{I}= b^\mathcal{I}$\\
individual inequality assertion & $a \not \approx b$ & $a^\mathcal{I} \not = b^\mathcal{I}$\\
\hline
\end{tabular}
\end{center}
\end{table*}

A DL knowledge base (KB) $\Sigma$ can state both is-a relations between concepts (\emph{axioms}) and instance-of
relations between individuals (resp. couples of individuals) and concepts (resp. roles) (\emph{assertions} or \emph{facts}). Axioms form the so-called \emph{terminological box} (TBox) $\mathcal{T}$ whereas facts are contained in the so-called \emph{assertional box} (ABox) $\mathcal{A}$. A $\mathcal{SHIQ}$ KB encompasses also a role box (RBox) $\mathcal{R}$ which consists of a finite set of role equivalence and role inclusion axioms. Therefore hierarchies can be defined over not only concepts but also roles. Transitivity of roles is also specified by means of axioms. Thus, when a DL-based ontology language is adopted, an ontology is nothing else than a TBox, possibly together with a RBox. If the ontology is populated, it corresponds to a whole DL KB, i.e. encompassing also an ABox.
The semantics of DLs can be defined directly with set-theoretic formalizations as shown in Table \ref{tab:DL-KB} or through a mapping to FOL as shown in \cite{Borgida96}. An \emph{interpretation}
$\mathcal{I}=(\Delta^\mathcal{I}, \cdot^\mathcal{I})$ for a DL KB consists of a domain $\Delta^\mathcal{I}$ and a mapping function
$\cdot^\mathcal{I}$. Under the \emph{Unique Names Assumption} (UNA)\cite{Reiter80b}, individuals are mapped to elements of
$\Delta^\mathcal{I}$ such that $a^\mathcal{I} \neq b^\mathcal{I}$ if $a \neq b$. Yet UNA does not hold by default in DLs. Thus individual equality (inequality) assertions may appear in a DL KB (see Table \ref{tab:DL-KB}). An interpretation $\mathcal{I}$ is a \emph{model} of a KB $\Sigma =(\mathcal{T}, \mathcal{A})$ iff it satisfies all axioms and assertions in $\mathcal{T}$ and $\mathcal{A}$ .
Also the KB represents many different interpretations, i.e. all its models. This is coherent with the \emph{Open World Assumption} (OWA) that holds in FOL semantics. A DL KB is \emph{satisfiable} if it has at least one model. An ABox assertion $\alpha$ is a \emph{logical consequence} of a KB $\Sigma$, written $\Sigma \models \alpha$, if all models of $\Sigma$ are also models of $\alpha$.

The main reasoning task for a DL KB $\Sigma$ is the \emph{consistency check} which tries to prove the satisfiability of $\Sigma$. The consistency check is performed by applying decision procedures mostly based on tableau calculus. Another well known reasoning service in DLs is \emph{instance check}, i.e., the check of whether an ABox assertion is a logical implication of a DL KB. A more sophisticated version of instance check, called \emph{instance retrieval}, retrieves, for a DL KB $\Sigma$, all (ABox) individuals that are instances of the given (possibly complex) concept expression $C$, i.e., all those individuals $a$ such that $\Sigma$ entails that $a$
is an instance of $C$. In data-intensive applications, querying KBs plays a central role. Instance retrieval
is, in some aspects, a rather weak form of querying: although possibly complex concept
expressions are used as queries, we can only query for tree-like relational structures, i.e.,
a DL concept cannot express arbitrary cyclic structures. 
The possibility of expressing \emph{conjunctive queries} (CQ) and \emph{unions of conjunctive queries} (UCQ) is widely studied in DLs.
Let $P_\mathcal{C}$ and $P_\mathcal{R}$ be the alphabets of concept names and role names, respectively. 
A \emph{Boolean UCQ} over the alphabet $P_\mathcal{C} \cup P_\mathcal{R}$ is a FOL sentence
of the form $q_1 \vee \ldots \vee q_n$, where each $q_i$ is a conjunction $\exists \vec{X} conj_i(\vec{X})$ of atoms
whose predicates are in $P_\mathcal{C} \cup P_\mathcal{R}$ and whose arguments are either
constants or variables from the tuple $\vec{X}$. A \emph{Boolean CQ} corresponds
to a Boolean UCQ in the case when $n = 1$. The \emph{Boolean UCQ entailment problem} in DLs is defined as follows: A KB $\Sigma$ entails a UCQ $Q = q_1 \vee \ldots \vee q_n$, written
as $\Sigma \models q_1 \vee \ldots \vee q_n$, if, for every model
$\mathcal{I}$ of $\Sigma$, there is some $i$ such that $q_i$ is satisfied in $\mathcal{I}$ and $1 \leq i \leq n$. Note that instance check can be expressed as the problem of query entailment problem of a Boolean CQs constituted by just one ground atom. The \emph{Boolean CQ/UCQ containment problem}\footnote{This problem was called \emph{existential entailment} in \cite{LevyR98}.} in DLs is defined as follows: Given a $\mathcal{DL}$-TBox $\mathcal{T}$, a Boolean CQ $Q_1$ and a Boolean
UCQ $Q_2$ over the alphabet $P_\mathcal{C} \cup P_\mathcal{R}$, $Q_1$ is contained in $Q_2$
with respect to $\mathcal{T}$, denoted by $\mathcal{T} \models Q_1 \subseteq Q_2$, iff, for every
model $\mathcal{I}$ of $\mathcal{T}$, if $Q_1$ is satisfied in $\mathcal{I}$ then $Q_2$ is satisfied in
$\mathcal{I}$. This problem has been proved decidable for many DLs, notably for the very expressive $\mathcal{SHIQ}$ \cite{GlimmHLS08} and $\mathcal{SHOQ}$ \cite{GlimmHS08-kr}. Finally, when the UNA does not hold, it can be immediately reduced to the Boolean UCQ entailment problem \cite{CalvaneseDL08}. In the rest of the paper we shall consider DLs without UNA.

\subsection{Integrating ontologies and relational databases}\label{sect:hybrid-kr}

The integration of ontologies and relational databases follows the tradition of KR research on \emph{hybrid systems}, i.e. those systems which are constituted by two or more subsystems dealing with distinct portions of a single KB by performing specific reasoning procedures \cite{FrischC91}. 
The motivation for investigating and developing such systems is to improve on two basic features of KR formalisms, namely \emph{representational adequacy} and \emph{deductive power}, by preserving the other crucial feature, i.e. \emph{decidability}. Those KR systems that integrate ontologies and relational databases will be referred to as DL-CL hybrid KR systems in the rest of the paper. They implement different solutions to the problem of combining DLs and CLs. Indeed DLs and CLs are FOL fragments incomparable as for the expressiveness \cite{Borgida96} and the semantics \cite{Rosati05} but combinable at different degrees of integration. The integration is said to be \emph{tight} when a model of the hybrid KB is defined as the union of two models, one for the DL part and one for the CL part, which share the same domain. In particular, combining DLs with CLs in a tight manner can easily yield to undecidability if the interaction scheme between the DL and the CL part of a hybrid KB does not fulfill some condition of \emph{safeness}\cite{Rosati05-ppswr}. Indeed safeness allows to solve the semantic mismatch between DLs and CLs, namely the OWA for DLs and the CWA for CLs\footnote{Note that the OWA and CWA have a strong influence on the results of reasoning.}. In the following we shall briefly describe two exemplary cases of tightly-integrated DL-CL hybrid KR systems: $\mathcal{AL}$-log \cite{Donini98} and \textsc{Carin} \cite{LevyR98}. The former is safe whereas the latter is not.

$\mathcal{AL}$-log \cite{Donini98} is a hybrid KR system that integrates $\mathcal{ALC}$ \cite{Schmidt-Schauss91} and \textsc{Datalog} \cite{CeriGT89}. In particular, variables occurring in the body of rules may be constrained with $\mathcal{ALC}$ concept assertions to be used as 'typing constraints'. This makes rules applicable only to explicitly named objects. A further restriction is that only \textsc{Datalog} atoms are allowed in rule heads. Reasoning for $\mathcal{AL}$-log knowledge bases is based on \emph{constrained
SLD-resolution}, i.e. an extension of SLD-resolution with a tableau calculus for $\mathcal{ALC}$ to deal with
constraints. Constrained SLD-resolution is \emph{decidable} and runs in single non-deterministic exponential time. Constrained SLD-refutation is a complete and sound method for answering \emph{ground} queries, i.e. conjunctions of ground \textsc{Datalog} atoms and $\mathcal{ALC}$ concept assertions. 

A comprehensive study of the effects of combining DLs and CLs can be found in \cite{LevyR98}. Here the family \textsc{Carin} of hybrid languages is presented. Special attention is devoted to the DL $\mathcal{ALCNR}$. The results of the study can be summarized as follows: (i) answering CQs over $\mathcal{ALCNR}$ TBoxes is decidable, (ii) query answering in a logic obtained by extending $\mathcal{ALCNR}$ with non-recursive \textsc{Datalog} rules, where both concepts and roles can occur in rule bodies, is also decidable, as it can be reduced to answering a UCQ, (iii) if rules are recursive, query answering becomes undecidable, (iv) decidability can be regained by disallowing certain combinations of constructors in the logic, and (v) decidability can be regained by requiring rules to be \emph{role-safe}, where at least one variable from each role literal must occur in some non-DL-atom. As in $\mathcal{AL}$-log, query answering is decided using constrained resolution and a modified version of tableau calculus.

\subsection{Learning rules with ILP}\label{sect:ilp}

Inductive Logic Programming (ILP) was born at the intersection between Logic Programming and Concept Learning \cite{Muggleton90}. From Logic Programming it has borrowed the KR framework, i.e. Horn Clausal Logic (HCL). From Concept Learning it has inherited the inferential mechanisms for induction, the most prominent of which is \emph{generalization}. Concept Learning is concerned with the problem of automatically inducing the general definition of some concept (called \emph{target}), given \emph{examples} labeled as instances or noninstances of the concept. In ILP the target is the predicate whose definition is returned by the inductive learning process as a \emph{hypothesis}. The definition may consist of one or more clauses. 
A distinguishing feature of ILP with respect to other forms of Concept Learning is the use of prior knowledge of the domain of interest, called \emph{background knowledge} (BK). Therefore, induction with ILP generalizes from individual instances/observations in the presence of BK, finding valid hypotheses. \emph{Validity} depends on the
underlying \emph{setting}.
 
\subsubsection{Settings}
At present, there exist several formalizations of
induction in ILP that can be classified according to the following two orthogonal dimensions: the \emph{scope of induction} (discrimination vs
characterization) and the \emph{representation of observations} (ground definite
clauses vs ground unit clauses) \cite{DeRaedtD97}. \emph{Discriminant induction} aims at
inducing hypotheses with discriminant power as required in tasks such as
classification where observations encompass both positive and
negative examples. \emph{Characteristic induction} is more suitable for finding regularities
in a data set. This corresponds to learning from positive
examples only. For a thorough discussion of
differences between discriminant and characteristic induction see \cite{Michalski83}.
The second dimension affects the notion of \emph{coverage}, i.e. the condition under which a hypothesis explains/confirms an observation. In \emph{learning from entailment} (also called normal or explanatory ILP setting), hypotheses are clausal theories, observations are ground definite clauses, and a
hypothesis covers an observation if the hypothesis logically entails the observation \cite{FrazierP93-icml}. In \emph{learning
from interpretations} (also called nonmonotonic or confirmatory ILP setting), hypotheses are clausal theories, observations are
Herbrand interpretations (ground unit clauses) and a hypothesis covers an observation if the observation is a model for the hypothesis \cite{DeRaedtD94}. Summing up, when learning from entailment with the aim of discrimination, a hypothesis is valid (or correct) if it logically entails all positive examples and none of the negative examples. The former condition of validity is called \emph{completeness}, whereas the latter is referred to as \emph{consistency}. If the scope of induction is characterization, the condition of consistency is dropped out from the notion of validity due to the absence of negative examples. The two settings for the case of learning from interpretations can be defined similarly.

\subsubsection{Techniques}
In Concept Learning, thus in ILP, generalization is traditionally viewed as search through a partially ordered space of inductive hypotheses \cite{Mitchell82}. According to this vision, an inductive hypothesis is a clausal theory and the induction of a single clause requires (i) structuring, (ii) searching and (iii) bounding the space of clauses \cite{Nienhuys97}.

First we focus on (i) by clarifying how the algebraic notion of ordering can be applied to clauses. A \emph{generality relation} allows for determining which one, between two clauses, is more general than the other. It defines a pre-order (or quasi order) on the set of clauses, i.e. a partially-ordered set of equivalence classes. One such ordering is \emph{$\theta$-subsumption} \cite{Plotkin70}: Given two clauses $C$ and $D$, we say that $C$ $\theta$-subsumes $D$ if there exists a substitution $\theta$, such that $C\theta \subseteq D$\footnote{This definition relies on the set notation for clauses.}. Given the usefulness of BK, orders have been proposed that reckon with it. Among them is \emph{relative subsumption} \cite{Plotkin71a}: Given two clauses $C$ and $D$ and a clausal theory $\mathcal{K}$, we say that $C$ subsumes $D$ relative to $\mathcal{K}$ if there exists a substitution $\theta$ such that $\mathcal{K} \models \forall (C\theta \implies D)$. Also, \emph{generalized subsumption} \cite{Buntine88} is of interest to this paper: Given two definite clauses $C$ and $D$ standardized apart\footnote{Two clauses $C$ and $D$ are said to be \emph{standardized apart} if they have no variables in common.} and a definite program $\mathcal{K}$, we say that $C$ subsumes $D$ w.r.t. $\mathcal{K}$ iff there
exists a ground substitution $\theta$ for $C$ such that (i)
$head(C)\theta=head(D)\sigma$ and (ii) $\mathcal{K} \cup body(D)\sigma \models
body(C)\theta$ where $\sigma$ is a Skolem substitution\footnote{Let $\mathcal{B}$ be a clausal theory and $C$ be a clause. Let $X_1,\ldots,X_n$ be all the variables appearing in $C$,
and $a_1,\ldots,a_n$ be distinct constants (individuals) not appearing in
$\mathcal{B}$ or $C$. Then the substitution $\{X_1/a_1,\ldots,X_n/a_n\}$ is called a
\emph{Skolem substitution} for $C$ w.r.t. $\mathcal{B}$.} 
for $D$ with respect to $\{C\} \cup \mathcal{K}$. 
In the general case, generalized subsumption is undecidable and does
not introduce a lattice\footnote{A \emph{lattice} is a partially ordered set (also called a poset) in which any two elements have a unique supremum (the elements' least upper bound) and an infimum (greatest lower bound).} on a set of clauses. Because of these problems, $\theta$-subsumption is more frequently used in ILP systems. Yet for \textsc{Datalog} generalized subsumption is decidable and admits a least general
generalization.

Once structured according to a generality order, the space of hypotheses can be searched (ii) by means of refinement operators. 
A \emph{refinement operator} is a function which computes a set of specializations or generalizations of a clause according to whether a top-down or a bottom-up search is performed. The two kinds of refinement operator have been therefore called \emph{downward} and \emph{upward}, respectively. A good refinement operator should satisfy certain desirable properties \cite{vanderLaag95}. We shall
illustrate these properties for the case of downward refinement operators but
analogous conditions are actually required to hold for the upward ones as
well. Ideally, a downward refinement operator should compute only a finite set
of specializations of each clause - otherwise it will be of limited practical use.
When it accomplishes this condition, it is called \emph{locally finite}. Furthermore, it should be
\emph{complete}: every specialization should be reachable by a finite number of
applications of the operator. Finally, it is better only to compute \emph{proper}
specializations of a clause, for otherwise repeated application of the operator might
get stuck in a sequence of equivalent clauses, without ever achieving any real
specialization. Operators that satisfy all these conditions simultaneously are called
\emph{ideal}. It has been shown that ideal refinement operators do
not exist for both full and Horn clausal languages ordered by either subsumption or
the stronger orders (e.g. implication). 

In order to define a refinement operator for
full clausal languages, it is necessary to drop one of the three properties of
idealness. Since local finiteness and completeness are usually considered the most
important among these properties, this means that locally finite and complete, but improper
refinement operators can be defined for full clausal languages. On the other hand, in
order to retain all the three properties of idealness, it seems that the only
possibility is to restrict the search space. Hence, the definition of refinement operators is usually coupled with the specification of a declarative bias for bounding the space of clauses (iii). \emph{Bias} concerns anything which
constrains the search for theories, e.g. a \emph{language bias} specifies syntactic constraints on the clauses in the search space. One such constraint is \emph{connectedness}: A clause $C$ is connected if each variable occurring in $head(C)$ also occurs in $body(C)$. The constraint of \emph{linkedness} is also widely used: A definite clause $C$ is linked if each literal $l_{i} \in C$ is linked. A literal $l_{i} \in C$ is
linked if at least one of its terms is linked. A term $t$ in some literal $l_{i}
\in C$ is linked with linking-chain of length 0, if $t$ occurs in $head(C)$,
and with linking-chain of length $d+1$, if some other term in $l_{i}$ is
linked with linking-chain of length $d$. The link-depth of a term $t$ in $l_{i}$ is the length of the shortest linking-chain of $t$.

\section{Integrating Ontologies and Databases with \textsc{$\mathcal{DL}$+log}$^{\neg\vee}$}\label{sect:dl+log}

The KR framework of \textsc{$\mathcal{DL}$+log}$^{\neg\vee}$ \cite{Rosati06} allows for the tight integration of DLs \cite{BaaderCMcGNPS07} and $\textsc{Datalog}^{\neg \vee}$ \cite{EiterGM97}. More precisely, it allows a $\mathcal{DL}$ KB to be extended with $\textsc{Datalog}^{\neg \vee}$ rules according to the so-called \emph{weak safeness} condition as shown in the following. 

\subsection{Syntax}

Formulas in \textsc{$\mathcal{DL}$+log}$^{\neg\vee}$ are built upon three mutually disjoint predicate alphabets: an alphabet $P_\mathcal{C}$ of concept names, an alphabet $P_\mathcal{R}$ of role names, and an alphabet $P_\textsc{D}$ of $\textsc{Datalog}$ predicates.
We call a predicate $p$ a \emph{DL-predicate} if either $p \in P_\mathcal{C}$ or
$p \in P_\mathcal{R}$. Then, we denote by $\mathcal{N}$ a countably infinite alphabet
of constant names. An \emph{atom} is an expression of the form $p(\vec{X})$, where $p$ is
a predicate of arity $n$ and $\vec{X}$ is a n-tuple of variables and
constants. If no variable symbol occurs in $\vec{X}$, then $p(\vec{X})$ is
called a \emph{ground atom} (or \emph{fact}). If $p \in P_\mathcal{C} \cup P_\mathcal{R}$, the atom
is called a \emph{DL-atom}, while if $p \in P_\textsc{D}$, it is called a \emph{$\textsc{Datalog}$
atom}.
\begin{definition}\label{def:dl+log-KB}
Given a description logic $\mathcal{DL}$, a \textsc{$\mathcal{DL}$+log}$^{\neg\vee}$ KB $\mathcal{B}$ is a pair $(\Sigma, \Pi)$, where $\Sigma$ is a $\mathcal{DL}$ KB and $\Pi$ is a set of $\textsc{Datalog}^{\neg \vee}$ rules, where each rule $R$ has the
form
\begin{center}
$p_1(\vec{X_1}) \vee \ldots \vee p_n(\vec{X_n}) \leftarrow$\\ $r_1(\vec{Y_1}), \ldots, r_m(\vec{Y_m}), s_1(\vec{Z_1}), \ldots, s_k(\vec{Z_k}), not$ $u_1(\vec{W_1}), \ldots, not$ $u_h(\vec{W_h})$ (1)
\end{center}
with $n, m, k, h \geq 0$, each $p_i(\vec{X_i})$, $r_j(\vec{Y_j})$, $s_l(\vec{Z_l})$,
$u_k(\vec{W_k})$ is an atom and:
\begin{itemize}
	\item each $p_i$ is either a DL-predicate or a $\textsc{Datalog}$ predicate;
	\item each $r_j$, $u_k$ is a $\textsc{Datalog}$ predicate;
	\item each $s_l$ is a DL-predicate;
	\item ($\textsc{Datalog}$-safeness) every variable occurring in
$R$ must appear in at least one of the atoms
$r_1(\vec{Y_1}), \ldots, r_m(\vec{Y_m}), s_1(\vec{Z_1}), \ldots, s_k(\vec{Z_k})$;
	\item (weak $\mathcal{DL}$-safeness) every head variable of $R$ must appear
in at least one of the atoms $r_1(\vec{Y_1}), \ldots, r_m(\vec{Y_m})$.
\end{itemize}
\end{definition}

We remark that the condition of weak $\mathcal{DL}$-safeness allows
for the presence of variables that only occur in DL-atoms
in the body of $R$. This condition allows to overcome the main representational limits of the safe approaches by keeping the integration scheme still decidable. Indeed, the notion of $\mathcal{DL}$-safeness proposed in \cite{MotikSS05}
can be expressed as follows: every variable of $R$ must appear in
at least one of the atoms $r_1(\vec{Y_1}), \ldots, r_m(\vec{Y_m})$. Therefore,
$\mathcal{DL}$-safeness forces every variable of $R$ to occur also in the
$\textsc{Datalog}$ atoms in the body of $R$. This disables the possibility of expressing CQs and UCQs. By weakening the $\mathcal{DL}$-safeness condition, this possibility can be enabled. For these reasons, \textsc{$\mathcal{DL}$+log}$^{\neg\vee}$ is located between $\mathcal{AL}$-log and \textsc{Carin} along the expressivity line.  

Without loss of generality, we can assume
that in a \textsc{$\mathcal{DL}$+log}$^{\neg\vee}$ KB $(\Sigma,\Pi)$ all constants occurring
in $\Sigma$ also occur in $\Pi$. 

\begin{example}\label{ex:dl+log-KB}
Let us consider a \textsc{$\mathcal{DL}$+log}$^{\neg\vee}$ KB $\mathcal{B}$ (adapted from \cite{Rosati06}) integrating the following DL-KB $\Sigma$ (ontology about persons)
\begin{tabbing}
  MM \= MMMMMM \= MM \= \kill
$[A1]$ \texttt{PERSON} $\sqsubseteq \exists$ \texttt{FATHER$^{-}$.MALE}\\
$[A2]$ \texttt{MALE} $\sqsubseteq$ \texttt{PERSON}\\
$[A3]$ \texttt{FEMALE} $\sqsubseteq$ \texttt{PERSON}\\
$[A4]$ \texttt{FEMALE} $\sqsubseteq \neg$\texttt{MALE}\\
\> \texttt{MALE(Bob)}\\
\> \texttt{PERSON(Mary)}\\
\> \texttt{PERSON(Paul)}\\
\> \texttt{FATHER(John,Paul)}
\end{tabbing}
and the following $\textsc{Datalog}^{\neg \vee}$ program $\Pi$ (database about students):
\begin{tabbing}
  MM \= MMMMMM \= MM \= \kill
$[R1]$ \texttt{boy(X)} $\leftarrow$ \texttt{enrolled(X,c1,ft)}, \texttt{PERSON(X)}, \texttt{not girl(X)}\\
$[R2]$ \texttt{girl(X)} $\leftarrow$ \texttt{enrolled(X,c2,ft)}, \texttt{PERSON(X)}\\
$[R3]$ \texttt{boy(X)}$\vee$ \texttt{girl(X)} $\leftarrow$ \texttt{enrolled(X,c3,ft)}, \texttt{PERSON(X)}\\
$[R4]$ \texttt{FEMALE(X)} $\leftarrow$ \texttt{girl(X)}\\
$[R5]$ \texttt{MALE(X)} $\leftarrow$ \texttt{boy(X)}\\
$[R6]$ \texttt{man(X)} $\leftarrow$ \texttt{enrolled(X,c3,pt)}, \texttt{FATHER(X,Y)}\\
\> \texttt{enrolled(Paul,c1,ft)}\\
\> \texttt{enrolled(Mary,c1,ft)}\\
\> \texttt{enrolled(Mary,c2,ft)}\\
\> \texttt{enrolled(Bob,c3,ft)}\\
\> \texttt{enrolled(John,c3,pt)}
\end{tabbing}
encompassing rules that mix DL-literals and $\textsc{Datalog}$-literals. The rule $[R3]$, e.g., says that: If \texttt{X} is a \texttt{PERSON} enrolled in the course \texttt{c3} as a full-time student (\texttt{ft}), then \texttt{X} is either a \texttt{boy} or a \texttt{girl}. The rule $[R6]$ says that: If \texttt{X} is a \texttt{FATHER} (of some \texttt{Y}) enrolled in the course \texttt{c3} as a part-time student (\texttt{pt}), then \texttt{X} is a \texttt{man}. Notice that the variable \texttt{Y} in $R6$ is weakly-safe but not DL-safe, since \texttt{Y} does not occur in any $\textsc{Datalog}$ literal of $R6$.
\end{example}

\subsection{Semantics}

For \textsc{$\mathcal{DL}$+log}$^{\neg\vee}$ two semantics have been defined: a FOL semantics and a NM semantics. The FOL semantics does not distinguish between head atoms and negated body atoms. Thus, the rule (1) is equivalent to:
\begin{center}
$p_1(\vec{X_1}) \vee \ldots \vee p_n(\vec{X_n}) \vee u_1(\vec{W_1})\vee \ldots \vee u_h(\vec{W_h}) \leftarrow$\\ $r_1(\vec{Y_1}), \ldots, r_m(\vec{Y_m}), s_1(\vec{Z_1}), \ldots, s_k(\vec{Z_k})$ (2)
\end{center}
The NM semantics is based on the stable model semantics of $\textsc{Datalog}^{\neg\vee}$. According to it, DL-predicates are still interpreted under OWA, while \textsc{Datalog} predicates are interpreted under CWA. Notice that, under both semantics, entailment can be reduced to satisfiability, since it is possible to express constraints in the $\textsc{Datalog}$ program. In particular, it is immediate to verify the following theorem on ground query answering \cite{Rosati06}. 
\begin{theorem}\label{thm:DL+log-ground-query-answering}
Given a \textsc{$\mathcal{DL}$+log}$^{\neg\vee}$ KB $(\Sigma,\Pi)$ and a ground atom $\alpha$, $(\Sigma,\Pi) \models \alpha$ iff $(\Sigma,\Pi\cup\{\leftarrow \alpha\})$ is unsatisfiable.
\end{theorem}
Analogously, CQ answering can be reduced to satisfiability in $\textsc{Datalog}^{\neg\vee}$, more precisely it can be performed by means of multiple satisfiability tests. Consequently, Rosati \shortcite{Rosati06} concentrates on the satisfiability problem in \textsc{$\mathcal{DL}$+log}$^{\neg\vee}$ KBs. It has been shown that, when the rules are made out of $\textsc{Datalog}^{\vee}$ (i.e., without negated atoms), the above two semantics are equivalent with respect to the satisfiability problem. In particular, FOL-satisfiability can always be reduced (in linear time) to NM-satisfiability by rewriting rules from the form (1) to the form (2). Hence, only the satisfiability problem under the NM semantics is deeply treated in \cite{Rosati06}.

\begin{example}\label{ex:dl+log-sem}
With reference to Example \ref{ex:dl+log-KB}, it can be easily verified that all NM-models for $\mathcal{B}$ satisfy
the following ground atoms:
\begin{enumerate}
	\item \texttt{boy(Paul)} (since rule $[R1]$ is always applicable for \texttt{\{X/Paul\}} and $[R1]$ acts like a default rule, which can be read as follows: if \texttt{X} is a person enrolled in course \texttt{c1}, then \texttt{X} is
a boy, unless we know for sure that \texttt{X} is a girl);
	\item \texttt{girl(Mary)} (since rule $[R2]$ is always applicable for \texttt{\{X/Mary\}});
	\item \texttt{boy(Bob)} (since rule $[R3]$ is always applicable for \texttt{\{X/Bob\}}, and, by rule $[R4]$, the conclusion \texttt{girl(Bob)} is inconsistent with $\Sigma$);
	\item \texttt{MALE(Paul)} (due to rule $[R5]$ and conclusion 1);
	\item \texttt{FEMALE(Mary)} (due to rule $[R4]$ and conclusion 2).
\end{enumerate}
Notice that $\mathcal{B} \models_{NM}$\texttt{FEMALE(Mary)}, while $\Sigma \not\models_{FOL}$
\texttt{FEMALE(Mary)}. In other words, adding rules has indeed an
effect on the conclusions one can draw about DL-predicates.
Moreover, such an effect also holds under the FOL semantics
of \textsc{$\mathcal{DL}$+log}-KBs, since it can be verified that $\mathcal{B} \models_{FOL}$\texttt{FEMALE(Mary)} in this case.
\end{example}

\subsection{Reasoning}

The problem statement of NM-satisfiability for finite \textsc{$\mathcal{DL}$+log}$^{\neg\vee}$ KBs relies on the aforementioned Boolean CQ/UCQ containment problem for the $\mathcal{DL}$ part and on the so-called DL-grounding of the \textsc{Datalog}$^{\neg\vee}$ component. In particular, DL-grounding is an adaptation of the grounding operation used in stable model semantics to the \textsc{$\mathcal{DL}$+log}$^{\neg\vee}$ case. 

Given a \textsc{$\mathcal{DL}$+log}$^{\neg\vee}$ KB $\mathcal{B}=(\Sigma, \Pi)$, we denote by $\mathcal{C}_\Pi$ the set of constants occurring in $\Pi$. The \emph{DL-grounding} of $\Pi$, denoted as $gr_p(\Pi)$, is a set of Boolean CQs obtained by grounding all and only the DL-parts of rule bodies and the DL-atoms appearing in rule heads in $\Pi$ with respect to the constants in $\mathcal{C}_\Pi$. Note that grounding in $gr_p(\Pi)$ is partial, since the variables that only occur in DL-atoms in the body of rules are not replaced by constants in $gr_p(\Pi)$. Similarly to $gr_p(\Pi)$, we define the partial grounding of $\Pi$ on $\mathcal{C}_\Pi$, denoted as $pgr(\Pi, \mathcal{C}_\Pi)$, as the program obtained from $\Pi$ by grounding with the constants in $\mathcal{C}_\Pi$ all variables except for the existential variables of rules that only occur in DL-atoms. Finally, given a partition $(G_P ,G_N)$ of $gr_p(\Pi)$, we denote by $\Pi(G_P , G_N)$ the ground $\textsc{Datalog}^{\neg \vee}$
program obtained from $pgr(\Pi, \mathcal{C}_\Pi)$ by taking into account the two sets $G_P$ and $G_N$ so that no DL-predicate occurs in such a program.

Let $G$ be a set of Boolean CQs. Then, we denote by $CQ(G)$ (resp. $UCQ(G)$) the Boolean CQ (resp. UCQ) corresponding to the conjunction (resp. disjunction) of all the Boolean CQs in $G$. The algorithm NMSAT-$\mathcal{DL}$+log for deciding NM-satisfiability of \textsc{$\mathcal{DL}$+log}$^{\neg\vee}$ KBs has a very simple structure (see Figure \ref{fig:nmsat}). It guesses a partition $(G_P ,G_N)$ of $gr_p(\Pi)$ that
is consistent with the $\mathcal{DL}$-KB $\Sigma = (\mathcal{T},\mathcal{A})$ (Boolean CQ/UCQ containment problem) and such that $\Pi(G_P , G_N)$ has a stable model. More details can be found in \cite{Rosati06}.

\begin{figure}
  \centering
\begin{tabbing}
    MMM \= MMM \= MMM \= MMM \= MMM \kill
    \textbf{NMSAT-$\mathcal{DL}$+\textsc{log}}($\mathcal{B}$)\\
		1.	satisfiable=false\\
    2.  \textbf{if there exists} a partition $(G_P,G_N)$ of $gr_p(\Pi)$ \textbf{such that}\\
		3.	\> (a) $\Pi(G_P,G_N)$ has a stable model \textbf{and}\\
		4.	\> (b) $\mathcal{T}\models CQ(\mathcal{A}\cup G_P)\subset UCQ(G_N)$\\
    5.  \> \textbf{then} satisfiable=true\\
    6. 	\textbf{endif}\\
    \textbf{return} satisfiable
\end{tabbing}
  \caption{The algorithm NMSAT-$\mathcal{DL}$+\textsc{log}}\label{fig:nmsat}
\end{figure}

The decidability of reasoning, thus of ground query answering, in \textsc{$\mathcal{DL}$+log}$^{\neg\vee}$ depends on the decidability of 
the Boolean CQ/UCQ containment problem in $\mathcal{DL}$. 
\begin{theorem}\label{thm:DL+log-decidability}
For any $\mathcal{DL}$, satisfiability
of \textsc{$\mathcal{DL}$+log}$^{\neg\vee}$ KBs (under both FOL and NM semantics) is decidable iff Boolean CQ/UCQ containment is
decidable in $\mathcal{DL}$ \cite{Rosati06}.
\end{theorem}
From Theorem \ref{thm:DL+log-decidability} and from previous results on
query answering and query containment in DLs, it follows the decidability of reasoning in several instantiations of \textsc{$\mathcal{DL}$+log}$^{\neg\vee}$. In all these decidable cases, ground queries can be answered by applying NMSAT-\textsc{$\mathcal{DL}$+log}. 

The complexity of reasoning in \textsc{$\mathcal{DL}$+log}$^{\neg\vee}$ depends on the specific $\mathcal{DL}$ chosen for instantiating the framework. We remind the reader to \cite{Rosati06} for the analysis of some cases.

\section{Inducing Database Views in \textsc{$\mathcal{DL}$+log}$^{\neg}$ with ILP}\label{sect:inducing-views}

In this section we consider the problem of defining a new view in a database whose schema is partly represented by an ontology. We suppose that there are tuples known to belong to the view as well as tuples known not to belong to the view. Cast in the \textsc{$\mathcal{DL}$+log}$^{\neg}$ framework, this problem boils down to the problem of building \textsc{$\mathcal{DL}$+log}$^{\neg}$ rules defining a \textsc{Datalog} predicate $p$ which stands for the view name. Tuples are ground \textsc{Datalog} facts that are true for $p$ if they belong to the view, false otherwise.
The database problem of interest can be reformulated as the following problem of discriminant induction.
\begin{definition}\label{def:learning-problem-1}
Given:
\begin{itemize}
	\item a \textsc{Datalog} database $\Pi$ and a $\mathcal{DL}$ ontology $\Sigma$ integrated into a \textsc{$\mathcal{DL}$+log}$^{\neg}$ KB  $\mathcal{B}$ (background theory);
	\item a \textsc{Datalog} predicate $p$ (target predicate);
	\item a set $\mathcal{O}$ of ground \textsc{Datalog} facts that are either true or false for $p$ (examples); and 
	\item a set $\mathcal{L}$ of constraints on the form of \textsc{$\mathcal{DL}$+log}$^{\neg}$ definitions for $p$ (language of hypotheses)
\end{itemize}
the problem of defining the view of name $p$ is to induce a set $\mathcal{H} \subset \mathcal{L}$ (hypothesis) of \textsc{$\mathcal{DL}$+log}$^{\neg}$ rules from $\mathcal{O}$ and $\mathcal{B}$ such that	$\mathcal{H}$ explains $\mathcal{O}$ by taking $\mathcal{B}$ into account.
\end{definition} 

We assume that the \emph{background theory} $\mathcal{B}$ in Definition \ref{def:learning-problem-1} is a \textsc{$\mathcal{DL}$+log}$^{\neg}$ KB which consists of an intensional part $\mathcal{K}$ (i.e., the TBox $\mathcal{T}$ plus the set $\Pi_R$ of rules) and an extensional part $\mathcal{F}$ (i.e., the ABox $\mathcal{A}$ plus the set $\Pi_F$ of facts). Also we denote by $P_\mathcal{C}(\mathcal{B})$, $P_\mathcal{R}(\mathcal{B})$, and $P_\textsc{D}(\mathcal{B})$ the sets of concept, role and \textsc{Datalog} predicate names occurring in $\mathcal{B}$, respectively. Note that $p \not\in P_\textsc{D}(\mathcal{B})$.

\begin{example}\label{ex:shiq+log-kb}
Throughout this section we shall consider a database $\Pi$ in the form of the following \textsc{Datalog}$^{\neg}$ program:
\begin{tabbing}
  MM \= MMMMMM \= MM \= \kill
\> \texttt{famous(Mary)}\\
\> \texttt{famous(Paul)}\\
\> \texttt{famous(Joe)}\\
\> \texttt{scientist(Joe)}
\end{tabbing}
containing also the rule 
\begin{tabbing}
  MM \= MMMMMM \= MM \= \kill
$[R1]$ \texttt{RICH(X)} $\leftarrow$ \texttt{famous(X)}, \texttt{not scientist(X)}
\end{tabbing}
linking the database to the ontology $\Sigma$ expressed as the following $\mathcal{DL}$ KB:
\begin{tabbing}
  MM \= MMMMMM \= MM \= \kill
$[A1]$ \texttt{RICH}$\sqcap$\texttt{UNMARRIED} $\sqsubseteq \exists$ \texttt{WANTS-TO-MARRY$^{-}$.$\top$}\\
$[A2]$ \texttt{WANTS-TO-MARRY}$\sqsubseteq$\texttt{LOVES}\\
\> \texttt{UNMARRIED(Mary)}\\
\> \texttt{UNMARRIED(Joe)}
\end{tabbing}
Note that $\Pi$ and $\Sigma$ can be integrated into a \textsc{$\mathcal{DL}$+log}$^{\neg}$ KB $\mathcal{B}$ (adapted from \cite{Rosati06}) that concerns the individuals \texttt{Mary}, \texttt{Joe}, and \texttt{Paul} and builds upon the alphabets $P_\mathcal{C}(\mathcal{B}) = \{\texttt{RICH/1}, \texttt{UNMARRIED/1}\}$, $P_\mathcal{R}(\mathcal{B}) = \{\texttt{WANTS-TO-MARRY/2}, \texttt{LOVES/2}\}$, and $P_\textsc{D}(\mathcal{B}) = \{\texttt{famous/1}, \texttt{scientist/1}\}$.
\end{example}


The \emph{language $\mathcal{L}$ of hypotheses} in Definition \ref{def:learning-problem-1} must allow for the generation of \textsc{$\mathcal{DL}$+log}$^{\neg}$ rules starting from three disjoint alphabets $P_\mathcal{C}(\mathcal{L}) \subseteq P_\mathcal{C}(\mathcal{B})$, $P_\mathcal{R}(\mathcal{L}) \subseteq P_\mathcal{R}(\mathcal{B})$, and $P_\textsc{D}(\mathcal{L}) \subseteq P_\textsc{D}(\mathcal{B})$. Also we distinguish between $P_\textsc{D}^{+}(\mathcal{L})$ and $P_\textsc{D}^{-}(\mathcal{L})$ in order to specify which $\textsc{Datalog}$ predicates can occur in positive and negative literals, respectively. More precisely, we consider \textsc{$\mathcal{DL}$+log}$^{\neg}$ rules of the form
\begin{center}
$p(\vec{X}) \leftarrow r_1(\vec{Y_1}), \ldots, r_m(\vec{Y_m}), s_1(\vec{Z_1}), \ldots, s_k(\vec{Z_k}), not$ $u_1(\vec{W_1}), \ldots, not$ $u_h(\vec{W_h})$ 
\end{center}
where the unique literal $p(\vec{X})$ in the head is formed out of a \textsc{Datalog}-predicate $p$ which represents the target predicate.
Note that the conditions of linkedness and connectedness usually assumed in ILP are guaranteed by the conditions of \textsc{Datalog} safeness and weak $\mathcal{DL}$ safeness valid in \textsc{$\mathcal{DL}$+log}$^{\neg\vee}$. 

\begin{example}\label{ex:hyp-lang1}
Suppose that the \textsc{Datalog}-predicate \texttt{happy} is the target and the set $P_\textsc{D}^{+}(\mathcal{L}^\texttt{happy}) \cup P_\mathcal{C}(\mathcal{L}^\texttt{happy}) \cup P_\mathcal{R}(\mathcal{L}^\texttt{happy})=\{\texttt{famous/1}, \texttt{RICH/1}, \texttt{LOVES/2},\texttt{WANTS-TO-MARRY/2}\}$ provides the building blocks for the language $\mathcal{L}^\texttt{happy}$. The following \textsc{$\mathcal{DL}$+log}$^{\neg}$ rules 
\begin{tabbing}
  MMMMMM \= MMMMMM \= MM \= \kill
	$R_1^\texttt{happy}$ \> \texttt{happy(X)} $\leftarrow$ \texttt{famous(X)}\\
	$R_2^\texttt{happy}$ \> \texttt{happy(X)} $\leftarrow$ \texttt{famous(X)}, \texttt{RICH(X)}\\
  $R_3^\texttt{happy}$ \> \texttt{happy(X)} $\leftarrow$ \texttt{famous(X)}, \texttt{LOVES(Y,X)}\\
  $R_4^\texttt{happy}$ \> \texttt{happy(X)} $\leftarrow$ \texttt{famous(X)}, \texttt{WANTS-TO-MARRY(Y,X)}
\end{tabbing}  
belonging to $\mathcal{L}^\texttt{happy}$ can be considered definitions for the target predicate \texttt{happy}. 
\end{example}

The set $\mathcal{O}$ of observations in Definition \ref{def:learning-problem-1} contains facts of the kind $p(\vec{a_i})$ where $p$ is the target predicate and $\vec{a_i}$ is a tuple of individuals occurring in the ABox $\mathcal{A}$. We assume $\mathcal{B} \cap \mathcal{O} = \emptyset$. Furthermore, the description of each observation $o_i \in \mathcal{O}$ is in the background theory and may be incomplete due to the inherent nature of \textsc{$\mathcal{DL}$+log}$^{\neg}$. Therefore, the normal ILP setting is the most appropriate to the learning problem in hand and can be extended to \textsc{$\mathcal{DL}$+log}$^{\neg}$ as follows. 
\begin{definition}\label{def:coverage_int1}
Let $R \in \mathcal{L}$ be a \textsc{$\mathcal{DL}$+log}$^{\neg}$ rule, $\mathcal{B}$ a \textsc{$\mathcal{DL}$+log}$^{\neg}$ KB, $p$ the target predicate, and $o_i = p(\vec{a_i}) \in \mathcal{O}$ a ground $\textsc{Datalog}$ fact. We say that \emph{$R$ covers $o_i$ under entailment w.r.t. $\mathcal{B}$} iff $\mathcal{B} \cup R \models p(\vec{a_i})$. 
\end{definition} 
Note that the coverage test can be reduced to query answering in \textsc{$\mathcal{DL}$+log}$^{\neg}$ KBs which in turn can be reformulated as a satisfiability problem of the KB.
\begin{example}\label{ex:coverage-test1}
The rule $R_4^\texttt{happy}$ mentioned in Example \ref{ex:hyp-lang1} covers the observation $o_\texttt{Mary} = \texttt{happy(Mary)}$ because $\mathcal{B} \cup R_4^\texttt{happy} \models \texttt{happy(Mary)}$. Indeed, all NM-models for $\mathcal{B}^\prime=\mathcal{B} \cup R_4^\texttt{happy}$ satisfy:
\begin{itemize}
	\item \texttt{famous(Mary)} is in $\mathcal{B}$;
	\item $\exists$ \texttt{WANTS-TO-MARRY}$^{-}$.$\top$\texttt{(Mary)}, due to the axiom $[A1]$ and to the fact that both \texttt{RICH(Mary)} and \texttt{UNMARRIED(Mary)} hold in every model of $\mathcal{B}^\prime$. In particular, \texttt{RICH(Mary)} holds because of $[R1]$;
	\item \texttt{happy(Mary)}, due to the above conclusions and to the
rule $R_4^\texttt{happy}$. Indeed, since $\exists$\texttt{WANTS-TO-MARRY}$^{-}$.$\top$\texttt{(Mary)}
holds in every model of $\mathcal{B}^\prime$, it follows that in
every model there exists a constant \texttt{x} such that
\texttt{WANTS-TO-MARRY(x,Mary)} holds in the model, consequently
from $R_4^\texttt{happy}$ it follows that \texttt{happy(Mary)} also
holds in the model.
\end{itemize}
Note that $R_4^\texttt{happy}$ does not cover the observations $o_\texttt{Joe}=\texttt{happy(Joe)}$ and $o_\texttt{Paul}=\texttt{happy(Paul)}$. More precisely, $\mathcal{B}^\prime \not\models \texttt{happy(Joe)}$ because \texttt{scientist(Joe)} holds in every model of $\mathcal{B}^\prime$, thus making the rule $[R1]$ not applicable for \texttt{\{X/Joe\}}, therefore  \texttt{RICH(Joe)} not derivable. Finally, $\mathcal{B}^\prime \not\models \texttt{happy(Paul)}$ because \texttt{UNMARRIED(Paul)} is not forced to hold in every model of $\mathcal{B}^\prime$, therefore $\exists$\texttt{WANTS-TO-MARRY}$^{-}$.$\top$\texttt{(Paul)} is not forced by $[A1]$ to hold in every such model.

It can be proved that also $R_3^\texttt{happy}$ covers only $o_\texttt{Mary}$, while $R_1^\texttt{happy}$ covers all the three observations and $R_2^\texttt{happy}$ covers $o_\texttt{Mary}$ and $o_\texttt{Paul}$ only.
\end{example}

In order to support the induction of \textsc{$\mathcal{DL}$+log}$^{\neg}$ rules with ILP techniques, the language $\mathcal{L}$ of hypotheses needs to be equipped with a generality order $\succeq$ so that $(\mathcal{L}, \succeq)$ is a search space. Therefore, the next two subsections, Section \ref{sect:hyp-space-1} and Section \ref{sect:ref-op-1}, are devoted to suggested techniques for structuring and searching the hypothesis space, respectively. Conversely, Section \ref{sect:foil-like-algo} sketches an ILP algorithm employing these techniques to solve the original problem of inducing database views.

\subsection{The hypothesis space}\label{sect:hyp-space-1}

The definition of a generality order for hypotheses in $\mathcal{L}$ must consider the peculiarities of \textsc{$\mathcal{DL}$+log}$^{\neg}$. One issue arises from the presence of NAF literals (i.e., negated \textsc{Datalog} literals) both in the background theory and in the language of hypotheses. As pointed out in \cite{Sakama01}, rules in normal logic programs are syntactically
regarded as Horn clauses by viewing the NAF-literal $\neg p(X)$ as an atom $not\_p(X)$
with the new predicate $not\_p$. Then any result obtained in ILP on Horn logic programs is directly
carried over to normal logic programs. Assuming one such treatment of NAF literals, we propose to adapt generalized subsumption \cite{Buntine88} to the case of \textsc{$\mathcal{DL}$+log}$^{\neg}$ rules and provide a characterization of the resulting generality order, denoted by $\succeq_{\mathcal{K}}^{\neg}$, that relies on the reasoning tasks known for \textsc{$\mathcal{DL}$+log}$^{\neg\vee}$ and from which a test procedure can be derived.   
\begin{definition}\label{def:K-subsumption-2}
Let $R_1, R_2 \in \mathcal{L}$ be two \textsc{$\mathcal{DL}$+log}$^{\neg}$ rules standardized apart, $\mathcal{K}$ a \textsc{$\mathcal{DL}$+log}$^{\neg}$ KB, and $\sigma$ a Skolem substitution for $R_2$ with
respect to $\{R_1\} \cup \mathcal{K}$. We say that $R_1$ is \emph{more general than} $R_2$ w.r.t. $\mathcal{K}$, denoted by $R_1 \succeq_{\mathcal{K}}^{\neg} R_2$, iff there
exists a ground substitution $\theta$ for $R_1$ such that (i)
$head(R_1)\theta=head(R_2)\sigma$ and (ii) $\mathcal{K} \cup body(R_2)\sigma \models
body(R_1)\theta$. We say that $R_1$ is \emph{strictly more general than} $R_2$ w.r.t. $\mathcal{K}$, denoted by $R_1 \succ_{\mathcal{K}}^{\neg} R_2$, iff $R_1 \succeq_{\mathcal{K}}^{\neg} R_2$ and $R_2 \not\succeq_{\mathcal{K}}^{\neg} R_1$. We say that $R_1$ is \emph{equivalent to} $R_2$ w.r.t. $\mathcal{K}$, denoted by $R_1 \equiv_{\mathcal{K}}^{\neg} R_2$, iff $R_1 \succeq_{\mathcal{K}}^{\neg} R_2$ and $R_2 \succeq_{\mathcal{K}}^{\neg} R_1$. 
\end{definition}

Note that condition (ii) is a variant of the Boolean CQ/UCQ containment problem because $body(R_2)\sigma$ and $body(R_1)\theta$ are both Boolean CQs. The difference between (ii) and the original formulation of the problem is that $\mathcal{K}$ encompasses not only a TBox but also a set of rules. Nonetheless this variant can be reduced to the satisfiability problem for finite \textsc{$\mathcal{DL}$+log}$^{\neg}$ KBs. Indeed the skolemization of $body(R_2)$ allows to reduce the Boolean CQ/UCQ containment problem to a CQ answering problem. Due to the aforementioned link between CQ answering and satisfiability, checking (ii) can be reformulated as proving that the KB $(\mathcal{T}, \Pi_R \cup body(R_2)\sigma \cup \{\leftarrow body(R_1)\theta\})$ is unsatisfiable. Once reformulated this way, (ii) can be solved by applying the algorithm NMSAT-\textsc{$\mathcal{DL}$+log}.
\begin{example}\label{ex:generality-test1}
Let us consider the hypotheses
\begin{tabbing}
  MMMMMM \= MMMMMM \= MM \= \kill
	$R_1^\texttt{happy}$ \> \texttt{happy(A)} $\leftarrow$ \texttt{famous(A)}\\
	$R_2^\texttt{happy}$ \> \texttt{happy(X)} $\leftarrow$ \texttt{famous(X)}, \texttt{RICH(X)}
\end{tabbing}
reported in Example \ref{ex:hyp-lang1} up to variable renaming. We want to check whether
\begin{center}
 $R_1^\texttt{happy} \succeq_{\mathcal{K}}^{\neg} R_2^\texttt{happy}$
\end{center}
holds. Let $\sigma =\{\texttt{X}/\texttt{a}\}$ be a Skolem substitution for $R_2^\texttt{happy}$ with respect to $\mathcal{K} \cup R_1^\texttt{happy}$ and $\theta =\{\texttt{A}/\texttt{a}\}$ a ground substitution for $R_1^\texttt{happy}$. Both conditions of Definition \ref{def:K-subsumption-2} are immediately verified. Thus, $R_1^\texttt{happy} \succeq_{\mathcal{K}}^{\neg} R_2^\texttt{happy}$. Since the viceversa does not hold, we can say that $R_1^\texttt{happy} \succ_{\mathcal{K}}^{\neg} R_2^\texttt{happy}$. Analogously, it can be proved that $R_1^\texttt{happy} \succ_{\mathcal{K}}^{\neg} R_3^\texttt{happy}$ and $R_1^\texttt{happy} \succ_{\mathcal{K}}^{\neg} R_4^\texttt{happy}$. Also, it turns out that $R_2^\texttt{happy}$ is incomparable under $\succeq_{\mathcal{K}}^{\neg}$ with $R_3^\texttt{happy}$ and $R_4^\texttt{happy}$. Finally, it can be proved that $R_3^\texttt{happy} \succ_{\mathcal{K}}^{\neg} R_4^\texttt{happy}$. In particular, the condition (ii)  $\mathcal{K} \cup \{\texttt{famous(a)},\texttt{LOVES(b,a)}\} \models \{\texttt{famous(a)},\texttt{WANTS-TO-MARRY(b,a)}\}$ is nothing else that a ground query answering problem in \textsc{$\mathcal{DL}$+log}$^{\neg}$. The entailment is guaranteed by the axiom $[A2]$. 
\end{example}

It can be proved that $\succ_{\mathcal{K}}^{\neg}$ is a decidable quasi-order (i.e. it is a reflexive and
transitive relation) for \textsc{$\mathcal{DL}$+log}$^{\neg}$ rules. In particular, the decidability of $\succ_{\mathcal{K}}^{\neg}$ follows from the decidability of \textsc{$\mathcal{DL}$+log}$^{\neg}$.

\subsection{A refinement operator}\label{sect:ref-op-1}

As pointed out in Section \ref{sect:hyp-space-1}, the space $(\mathcal{L}, \succ_{\mathcal{K}}^{\neg})$ is a quasi-ordered set, therefore it can be searched by refinement operators. In the following, we define a downward refinement operator for a \textsc{$\mathcal{DL}$+log}$^{\neg}$ language.

\begin{definition}\label{def:ref-op-1}
Let $\mathcal{L}$ be a \textsc{$\mathcal{DL}$+log}$^{\neg}$ language of hypotheses built out of the three finite and disjoint alphabets $P_\mathcal{C}(\mathcal{L})$, $P_\mathcal{R}(\mathcal{L})$, and $P_\textsc{D}^{+}(\mathcal{L}) \cup P_\textsc{D}^{-}(\mathcal{L})$, and
\begin{center}
$p(\vec{X}) \leftarrow r_1(\vec{Y_1}), \ldots, r_m(\vec{Y_m}), s_1(\vec{Z_1}), \ldots, s_k(\vec{Z_k}), not$ $u_1(\vec{W_1}), \ldots, not$ $u_h(\vec{W_h})$
\end{center}
be a rule $R$ belonging to $\mathcal{L}$.
We define a \emph{downward refinement operator} $\rho^{\neg}$ for
$(\mathcal{L}, \succeq_{\mathcal{K}})$ such that the set
$\rho^{\neg}(R)$ contains all $R^\prime \in\mathcal{L}$ that can be obtained from $R$ by
applying one of the following refinement rules:
\begin{description}
  \item [$\langle AddDataLit\_B^{+} \rangle$] $body(R^\prime)=body(R) \cup \{r_{m+1}(\vec{Y_{m+1}})\}$ if
		\begin{enumerate}
			\item $r_{m+1} \in P_\textsc{D}^{+}(\mathcal{L})$
			\item $r_{m+1}(\vec{Y_{m+1}})\not\in body(R)$
		\end{enumerate}
\end{description}
\begin{description}
  \item [$\langle AddDataLit\_B^{-} \rangle$] $body(R^\prime)=body(R) \cup \{$$not$ $u_{m+1}(\vec{W_{h+1}})\}$ if
		\begin{enumerate}
			\item $u_{h+1} \in P_\textsc{D}^{-}(\mathcal{L})$
			\item $u_{h+1}(\vec{W_{h+1}})\not\in body(R)$
		\end{enumerate}
\end{description}
\begin{description}
  \item [$\langle AddOntoLit\_B \rangle$] $body(R^\prime)=body(R) \cup \{s_{k+1}(\vec{Z_{k+1}})\}$ if
  	\begin{enumerate}
			\item $s_{k+1} \in P_\mathcal{C}(\mathcal{L}) \cup P_\mathcal{R}(\mathcal{L})$
			\item it does not exist any $s_{l}\in body(H)$ such that $s_{k+1} \sqsubseteq s_{l}$
		\end{enumerate}
\end{description}
\begin{description}
  \item [$\langle SpecOntoLit\_B \rangle$] $body(R^\prime)= (body(R) \setminus \{s_{l}(\vec{Z_{l}})\}) \cup s_{l}^\prime(\vec{Z_{l}})$ if
  	\begin{enumerate}
			\item $s_{l}^\prime \in P_\mathcal{C}(\mathcal{L}) \cup P_\mathcal{R}(\mathcal{L})$
			\item $s_{l}^\prime \sqsubseteq s_{l}$
		\end{enumerate}
\end{description}
\end{definition}

All the rules of $\rho^{\neg}$ are correct, i.e. the $R^\prime$'s obtained by applying any of the
rules of $\rho^{\neg}$ to $R\in \mathcal{L}$ are such that $R \succ_{\mathcal{K}}^{\neg} R^\prime$. This can be
proved intuitively by observing that they act only on $body(R)$. Thus condition (i) of
Definition \ref{def:K-subsumption-2} is satisfied. Furthermore, it is straightforward
to notice that the application of any of the
rules of $\rho^{\neg}$ to $R$ reduces the number of
models of $R$. In particular, as for $\langle SpecOntoLit \rangle$, this
intuition follows from the semantics of DLs. So condition (ii) also is fulfilled.
\begin{example}\label{ex:ref-op-1}
With reference to Example \ref{ex:hyp-lang1}, applying the refinement rule $\langle AddDataLit\_B^{+} \rangle$ to
\begin{tabbing}
  MMMMMM \= MMMMMM \= MM \= \kill
	$R_0^\texttt{happy}$ \> \texttt{happy(X)} $\leftarrow$
\end{tabbing}  
produces $R_1^\texttt{happy}$ which can be further specialized into $R_2^\texttt{happy}$, $R_3^\texttt{happy}$, and $R_4^\texttt{happy}$ by means of $\langle AddOntoLit\_B \rangle$. 
Note that no other refinement rule can be applied to $R_1^\texttt{happy}$ and that $R_4^\texttt{happy}$ can be also obtained as refinement via $\langle SpecOntoLit\_B \rangle$ from $R_3^\texttt{happy}$. 
\end{example}

Ideal refinement operators have been proven not to exist
for clausal languages ordered by $\theta$-subsumption or stronger
orders but can be approximated by dropping the requirement of properness
or by bounding the language \cite{Nienhuys97}. We choose the latter option because it guarantees
that, if $(\mathcal{L}, \succeq)$ is a quasi-ordered set, $\mathcal{L}$ is finite and
$\succeq$ is decidable, then there always exists an ideal refinement operator for
$(\mathcal{L}, \succeq)$. In our case, since $\succ_{\mathcal{K}}^{\neg}$ is a decidable
quasi-order for any $\mathcal{DL}$ with decidable Boolean CQ/UCQ containment problem, we only need to bound $\mathcal{L}$ in a suitable manner. From Definition \ref{def:ref-op-1} we know that the alphabets $P_\mathcal{C}(\mathcal{L})$, $P_\mathcal{R}(\mathcal{L})$, and $P_\textsc{D}^{+}(\mathcal{L}) \cup P_\textsc{D}^{-}(\mathcal{L})$ are finite. Having \textsc{Datalog} as basis for the CL part of \textsc{$\mathcal{DL}$+log}$^{\neg}$ avoids the generation of infinite terms. Yet, the expressive power of \textsc{$\mathcal{DL}$+log}$^{\neg}$ requires several other bounds to be imposed on $\mathcal{L}$ in order to guarantee its finiteness. It is necessary to introduce a complexity measure for
\textsc{$\mathcal{DL}$+log}$^{\neg}$ rules, as a pair of two different coordinates. Considering that the
complexity of a \textsc{$\mathcal{DL}$+log}$^{\neg}$ rule resides in its body, the former coordinate is the size (i.e. the difference between the number of symbol occurrences and the number of distinct variables) of the biggest literal in
$body(R)$, while the latter is the number of literals in $body(R)$. To keep $\mathcal{L}$ finite, we need first to set a maximum value for these two coordinates. Second, it is necessary to set the maximum number of specialization/generalization steps of the DL literals so that the search in the ontology is also depth-bounded.

\subsection{An algorithm}\label{sect:foil-like-algo}

\begin{figure}[t]
  \centering
\begin{tabbing}
    MMM \= MMM \= MMM \= MMM \= MMM \kill
    NMLEARN-$\mathcal{DL}$+\textsc{log}$^{\neg}$($\mathcal{L}$, $\mathcal{B}$, $\mathcal{O}$, $p$)\\
    1.  $\mathcal{H}\leftarrow \emptyset$\\
    2.	$E^{+} \leftarrow \{o_i \in \mathcal{O} | o_i \ is \ true \ for \ p\}$;\\
    3.	$E^{-} \leftarrow \{o_i \in \mathcal{O} | o_i \ is \ false \ for \ p\}$;\\
    4.  \textbf{while} $E^{+} \neq \emptyset$ \textbf{do}\\
		5.	\> $R \leftarrow \{ p(\vec{X}) \leftarrow \}$;\\
		6.	\> $E^{-}_R \leftarrow E^{-}$\\
    7.  \> \textbf{while} $E^{-}_R \neq \emptyset$ \textbf{do}\\
    8.  \> \> $\mathcal{Q} \leftarrow \{ R^\prime \in \mathcal{L}| R^\prime \in \rho^{\neg}(R)\}$;\\
		9.	\> \> $R \leftarrow best\_of(\mathcal{Q})$;\\
    10.	\> \> $E^{-}_R \leftarrow E^{-}_R \setminus \{ e \in E^{-}_R | \mathcal{B} \cup R \models e\}$;\\
    11.  \> \textbf{endwhile}\\
    12.	\> $\mathcal{H} \leftarrow \mathcal{H} \cup \{ R \}$;\\
    13. \> $E^{+} \leftarrow E^{+} \setminus \{ e \in E^{+} | \mathcal{B} \cup R \models e \}$;\\	
    14. \textbf{endwhile}\\
    \textbf{return} $\mathcal{H}$
\end{tabbing}
  \caption{Main procedure of NMLEARN-$\mathcal{DL}$+\textsc{log}$^{\neg}$}\label{fig:foil-like-algo}
\end{figure}

The algorithm in Figure \ref{fig:foil-like-algo} defines the main procedure of NMLEARN-$\mathcal{DL}$+\textsc{log}$^{\neg}$. Notice that the outer loop (4-14) corresponds to a variant of the sequential covering algorithm, i.e., it learns new rules one at a time, removing the positive examples covered by the latest rule before attempting to learn the next rule (13). The hypothesis space search performed by NMLEARN-$\mathcal{DL}$+\textsc{log}$^{\neg}$ is best understood by viewing it hierarchically. Each iteration through the outer loop (4-14) adds a new rule to its disjunctive hypothesis $\mathcal{H}$. The effect of each new rule is to generate the current disjunctive hypothesis (i.e., to increase the number of instances it classifies as positive), by adding a new disjunct. Viewed at this level, the search is a bottom-up search through the space of hypotheses, beginning with the most specific empty disjunction (1) and terminating when the hypothesis is sufficiently general to cover all positive training examples (14). The inner loop (7-11) performs a finer-grained search to determine the exact definition of each new rule. This loop searches a second hypothesis space, consisting of conjunctions of literals, to find a conjunction that will form the preconditions for the new rule. Within this space, it conducts a top-down, hill-climbing search, beginning with the most general preconditions possible (5), then adding literals one at a time to specialize the rule (7) until it avoids all negative examples. To select the most promising specialization from the candidates generated at each step (9), NMLEARN-$\mathcal{DL}$+\textsc{log}$^{\neg}$ considers the performance of the rule over the training examples, i.e. it maximizes the number of positive examples covered while keeping the number of negative examples covered as low as possible. 

\begin{example}\label{ex:foil-like-algo}
With reference to Example \ref{ex:ref-op-1} and Example \ref{ex:coverage-test1}, we suppose that 
\begin{tabbing}
  MM \= MMMMMM \= MM \= \kill
\> $E^{+} = \{o_\texttt{Mary}, o_\texttt{Joe}\}$\\
\> $E^{-} = \{o_\texttt{Paul}\}$
\end{tabbing}
The outer loop of the algorithm NMLEARN-$\mathcal{DL}$+\textsc{log}$^{\neg}$ starts from:
\begin{tabbing}
	MMMMMM \= MMMMMM \= MM \= \kill
	$R_0^\texttt{happy}$ \> \texttt{happy(X)} $\leftarrow$
\end{tabbing}
which is further refined through the iterations of the inner loop, more precisely it is first specialized into:
\begin{tabbing}
  MMMMMM \= MMMMMM \= MM \= \kill
	$R_1^\texttt{happy}$ \> \texttt{happy(X)} $\leftarrow$ \texttt{famous(X)}
\end{tabbing}
which in turn, since it covers negative examples, is then specialized into:
\begin{tabbing}
  MMMMMM \= MMMMMM \= MM \= \kill
	$R_2^\texttt{happy}$ \> \texttt{happy(X)} $\leftarrow$ \texttt{famous(X)}, \texttt{RICH(X)}\\
  $R_3^\texttt{happy}$ \> \texttt{happy(X)} $\leftarrow$ \texttt{famous(X)}, \texttt{LOVES(Y,X)}\\
  $R_4^\texttt{happy}$ \> \texttt{happy(X)} $\leftarrow$ \texttt{famous(X)}, \texttt{WANTS-TO-MARRY(Y,X)}
\end{tabbing}
out of which the rule $R_3^\texttt{happy}$ is selected as the best and added to the hypothesis because it does not cover negative examples. Note that $R_3^\texttt{happy}$ is preferred to $R_4^\texttt{happy}$ because it is more general. 
\end{example}

\section{Inducing Database Constraints in \textsc{$\mathcal{DL}$+log}$^{\neg\vee}$ with ILP}\label{sect:inducing-constraints}

In this section we face the problem of inducing an integrity theory $\mathcal{H}$ for a database $\Pi$ whose instance $\Pi_F$ is given and whose schema $\mathcal{K}$ encompasses an ontology $\Sigma$ and a set $\Pi_R$ of rules linking the database to the ontology. We assume that $\Pi$ and $\Sigma$ shares a common set of constants so that they can constitute a \textsc{$\mathcal{DL}$+log}$^{\neg\vee}$ KB $\mathcal{B}$.

\begin{definition}\label{def:learning-problem-2}
Given:
\begin{itemize}
	\item an intensional \textsc{Datalog} database $\Pi_R$ and a $\mathcal{DL}$ ontology $\Sigma$ integrated into a \textsc{$\mathcal{DL}$+log}$^{\neg\vee}$ KB  $\mathcal{K}$ (background theory);
	\item a set $\mathcal{O} = \Pi_F$ of ground \textsc{Datalog} facts (observation); and 
	\item a set $\mathcal{L}$ of constraints on the form of \textsc{$\mathcal{DL}$+log}$^{\neg\vee}$ rules to be induced (language of hypotheses)
\end{itemize}
the problem of defining an integrity theory for $\Pi_F$ is to induce a set $\mathcal{H} \subset \mathcal{L}$ (hypothesis) of \textsc{$\mathcal{DL}$+log}$^{\neg\vee}$ rules from $\mathcal{O}$ and $\mathcal{K}$ such that $\mathcal{H}$ confirms $\mathcal{O}$ by taking $\mathcal{K}$ into account.
\end{definition} 

Note that, as opposite to the learning problem formally stated in Definition \ref{def:learning-problem-1}, the \emph{background theory} in Definition \ref{def:learning-problem-2} is a \textsc{$\mathcal{DL}$+log}$^{\neg\vee}$ KB $\mathcal{K}$ which does not include the extensional part $\Pi_F$ of the database. Indeed $\Pi_F$ plays the role of the unique \emph{observation} from which the learning process should induce a theory $\mathcal{H}$. Conversely, similarly to Section \ref{sect:inducing-views}, we denote by $P_\mathcal{C}(\mathcal{B})$, $P_\mathcal{R}(\mathcal{B})$, and $P_\textsc{D}(\mathcal{B})$ the sets of concept, role and \textsc{Datalog} predicate names occurring in $\mathcal{B}$, respectively, assuming that $\mathcal{B} = \Sigma \cup \Pi$.

\begin{example}\label{ex:dl+log-kb}
Throughout this section we shall refer to a database about students in the form of a $\textsc{Datalog}^{\neg \vee}$ program $\Pi$ which consists of an extensional part $\Pi_F$ with the following facts:
\begin{tabbing}
  MM \= MMMMMM \= MM \= \kill
\> \texttt{boy(Paul)}\\
\> \texttt{girl(Mary)}\\
\> \texttt{enrolled(Paul,c1)}\\
\> \texttt{enrolled(Mary,c1)}\\
\> \texttt{enrolled(Mary,c2)}\\
\> \texttt{enrolled(Bob,c3)}
\end{tabbing}
and an intensional part $\Pi_R$ with the following rules: 
\begin{tabbing}
  MM \= MMMMMM \= MM \= \kill
$[R1]$  \texttt{FEMALE(X)} $\leftarrow$ \texttt{girl(X)}\\
$[R2]$  \texttt{MALE(X)} $\leftarrow$ \texttt{boy(X)}
\end{tabbing}
linking the database to an ontology about persons expressed as the following $\mathcal{DL}$ KB $\Sigma$:
\begin{tabbing}
  MM \= MMMMMM \= MM \= \kill
$[A1]$ \texttt{PERSON} $\sqsubseteq \exists$ \texttt{FATHER$^{-}$.MALE}\\
$[A2]$ \texttt{MALE} $\sqsubseteq$ \texttt{PERSON}\\
$[A3]$ \texttt{FEMALE} $\sqsubseteq$ \texttt{PERSON}\\
$[A4]$ \texttt{FEMALE} $\sqsubseteq \neg$\texttt{MALE}\\
\> \texttt{MALE(Bob)}\\
\> \texttt{PERSON(Mary)}\\
\> \texttt{PERSON(Paul)}
\end{tabbing}
Note that $\Pi$ and $\Sigma$ can be integrated into a \textsc{$\mathcal{DL}$+log}$^{\neg\vee}$ KB $\mathcal{B}$ (adapted from \cite{Rosati06}) that concerns the individuals \texttt{Bob}, \texttt{Mary}, and \texttt{Paul} and builds upon the alphabets $P_\mathcal{C}(\mathcal{B}) = \{\texttt{FEMALE/1}, \texttt{MALE/1}, \texttt{PERSON/1}\}$, $P_\mathcal{R}(\mathcal{B}) = \{\texttt{FATHER/2}\}$, and $P_\textsc{D}(\mathcal{B}) = \{\texttt{boy/1}, \texttt{girl/1}, \texttt{enrolled/2}\}$. 
\end{example}

The \emph{language $\mathcal{L}$ of hypotheses} in Definition \ref{def:learning-problem-2} must allow for the generation of \textsc{$\mathcal{DL}$+log}$^{\neg\vee}$ rules starting from three disjoint alphabets $P_\mathcal{C}(\mathcal{L}) \subseteq P_\mathcal{C}(\mathcal{B})$, $P_\mathcal{R}(\mathcal{L}) \subseteq P_\mathcal{R}(\mathcal{B})$, and $P_\textsc{D}(\mathcal{L}) \subseteq P_\textsc{D}(\mathcal{B})$. Analogously to Section \ref{sect:inducing-views}, we distinguish between $P_\textsc{D}^{+}(\mathcal{L})$ and $P_\textsc{D}^{-}(\mathcal{L})$ in order to specify which $\textsc{Datalog}$ predicates can occur in positive and negative literals, respectively. 

\begin{example}\label{ex:hyp-lang-2}
The following \textsc{$\mathcal{DL}$+log}$^{\neg\vee}$ rules: 
\begin{tabbing}
  MM \= MMMMMM \= MM \= \kill
\> \texttt{PERSON(X)} $\leftarrow$ \texttt{enrolled(X,c1)}\\
\> \texttt{boy(X)} $\vee$ \texttt{girl(X)} $\leftarrow$ \texttt{enrolled(X,c1)}\\
\> $\leftarrow$ \texttt{enrolled(X,c2)}, \texttt{MALE(X)}\\
\> $\leftarrow$ \texttt{enrolled(X,c2)}, \texttt{not girl(X)}\\
\> \texttt{MALE(X)} $\leftarrow$ \texttt{enrolled(X,c3)}
\end{tabbing}
belong to the language $\mathcal{L}$ built upon the alphabets $P_\mathcal{C}(\mathcal{L}) = P_\mathcal{C}(\mathcal{B})$, $P_\mathcal{R}(\mathcal{L}) = \emptyset$, $P_\textsc{D}^{+}(\mathcal{L})$ = \{\texttt{boy/1}, \texttt{girl/1}, \texttt{enrolled(\_,c1)}, \texttt{enrolled(\_,c2)}, \texttt{enrolled(\_,c3)}\}, and $P_\textsc{D}^{-}(\mathcal{L})$ = \{\texttt{boy/1}, \texttt{girl/1}\}.
\end{example}

The scope of induction in the learning problem of interest is characterization because we are looking for a theory which confirms the observation. Also, since a \textsc{$\mathcal{DL}$+log}$^{\neg\vee}$ KB may be incomplete due to the inherent nature of this KR framework, the most appropriate setting for induction is the one for learning from entailment. The coverage test proposed in the following generalizes the case illustrated in Definition \ref{def:coverage_int1} to observations which are not singletons of facts.
\begin{definition}\label{def:coverage_int2}
Let $R \in \mathcal{L}$ be a \textsc{$\mathcal{DL}$+log}$^{\neg\vee}$ rule, $\mathcal{K}$ a \textsc{$\mathcal{DL}$+log}$^{\neg\vee}$ KB, and $\mathcal{O} = \{ p_i(\vec{a_i}) \}$ a set of ground $\textsc{Datalog}$ facts. We say that \emph{$R$ covers $\mathcal{O}$ under entailment w.r.t. $\mathcal{K}$} iff $\mathcal{K} \cup R \models \bigwedge p_i(\vec{a_i})$. 
\end{definition} 
It is immediate to notice that the coverage test of Definition \ref{def:coverage_int2} can be reduced to Boolean CQ answering in \textsc{$\mathcal{DL}$+log}$^{\neg\vee}$ KBs and therefore to a NM-satisfiability problem.

In the following we sketch the ingredients for an ILP system able to discover such integrity theories on the basis of NMSAT-$\mathcal{DL}$+\textsc{log}. 

\subsection{The hypothesis space}\label{sect:hyp-space-2}

The order of relative subsumption \cite{Plotkin71a} is suitable for extension to \textsc{$\mathcal{DL}$+log}$^{\neg\vee}$ rules because it can cope with arbitrary clauses and admit an arbitrary finite set of clauses as the background theory. 
\begin{definition}\label{def:hyp-order-2}
Let $R_1, R_2 \in \mathcal{L}$ be two \textsc{$\mathcal{DL}$+log}$^{\neg\vee}$ rules, and $\mathcal{K}$ a \textsc{$\mathcal{DL}$+log}$^{\neg\vee}$ KB. We say that $R_1$ \emph{is more general than} $R_2$ w.r.t. $\mathcal{K}$, denoted by $R_1 \succeq_{\mathcal{K}}^{\neg\vee} R_2$, if there exists a substitution $\theta$ such that $\mathcal{K} \models \forall (R_1 \theta \implies R_2)$. We say that $R_1$ is \emph{strictly more general than} $R_2$ w.r.t. $\mathcal{K}$, denoted by $R_1 \succ_{\mathcal{K}}^{\neg\vee} R_2$, iff $R_1 \succeq_{\mathcal{K}}^{\neg\vee} R_2$ and $R_2 \not\succeq_{\mathcal{K}}^{\neg\vee} R_1$. We say that $R_1$ is \emph{equivalent to} $R_2$ w.r.t. $\mathcal{K}$, denoted by $R_1 \equiv_{\mathcal{K}}^{\neg\vee} R_2$, iff $R_1 \succeq_{\mathcal{K}}^{\neg\vee} R_2$ and $R_2 \succeq_{\mathcal{K}}^{\neg\vee} R_1$. 
\end{definition}

\begin{example}\label{ex:hyp-order-2}
Let us consider the following \textsc{$\mathcal{DL}$+log}$^{\neg\vee}$ rules belonging to the language $\mathcal{L}$ specified in Example \ref{ex:hyp-lang-2}:
\begin{tabbing}
  MMMM \= MMMMMM \= MM \= \kill
$R_1$ \> \texttt{boy(X)} $\leftarrow$ \texttt{enrolled(X,c1)}\\ 
$R_2$ \> \texttt{boy(A)} $\vee$ \texttt{girl(A)} $\leftarrow$ \texttt{enrolled(A,c1)}
\end{tabbing}
It can be easily proved that $R_1 \succeq_{\mathcal{K}}^{\neg\vee} R_2$. Let $\theta = \{\texttt{X}/\texttt{A}\}$ be the substitution to be applied to $R_1$ and let us suppose that, for every \texttt{A}, if \texttt{A} is enrolled in the course \texttt{c1}, then \texttt{A} is a \texttt{boy} (i.e. the rule $R_1 \theta$ is true), thus we can also say that \texttt{A} is either a \texttt{boy} or a \texttt{girl} (i.e. the rule $R_2$ is true). Note that $R_2 \not\succeq_{\mathcal{K}}^{\neg\vee} R_1$.

Let us now consider the following \textsc{$\mathcal{DL}$+log}$^{\neg\vee}$ rules also belonging to $\mathcal{L}$:
\begin{tabbing}
  MMMM \= MMMMMM \= MM \= \kill
$R_3$ \> \texttt{MALE(X)} $\leftarrow$ \texttt{enrolled(X,c1)}\\ 
$R_4$ \> \texttt{PERSON(A)} $\leftarrow$ \texttt{enrolled(A,c1)}
\end{tabbing}
In order to prove that $R_3 \succeq_{\mathcal{K}}^{\neg\vee} R_4$, we apply $\theta = \{\texttt{X}/\texttt{A}\}$ to $R_3$ and suppose that, for every \texttt{A}, if \texttt{A} is enrolled in the course \texttt{c1}, then \texttt{A} is a \texttt{MALE} (i.e. the rule $R_1 \theta$ is true). Due to axiom $[A2]$ occurring in the ontology $\Sigma$ reported in Example \ref{ex:dl+log-kb}, \texttt{A} is a \texttt{PERSON} (i.e. the rule $R_4$ is true). It is immediate to verify that $R_3 \succ_{\mathcal{K}}^{\neg\vee} R_4$.
\end{example}

The generality relation defined by $\succeq_{\mathcal{K}}^{\neg\vee}$ is a quasi-order on \textsc{$\mathcal{DL}$+log}$^{\neg\vee}$ rules, therefore the resulting space $(\mathcal{L}, \succeq_{\mathcal{K}}^{\neg\vee})$ can be searched by means of refinement operators.

\subsection{The refinement operator}\label{sect:ref-op-2}

A refinement operator for $(\mathcal{L}, \succeq_{\mathcal{K}}^{\neg\vee})$ should generate \textsc{$\mathcal{DL}$+log}$^{\neg\vee}$ rules good at expressing integrity constraints. Since we assume the database $\Pi$ and the ontology $\Sigma$ to be correct, a rule $R$ must be modified to make it satisfiable by $\Pi \cup \Sigma$ by either (i) strenghtening $body(R)$ or (ii) weakening $head(R)$.
\begin{definition}\label{def:ref-op-2}
Let $\mathcal{L}$ be a \textsc{$\mathcal{DL}$+log}$^{\neg\vee}$ language of hypotheses built out of the three finite and disjoint alphabets $P_\mathcal{C}(\mathcal{L})$, $P_\mathcal{R}(\mathcal{L})$, and $P_\textsc{D}^{+}(\mathcal{L}) \cup P_\textsc{D}^{-}(\mathcal{L})$, and
\begin{center}
$p_1(\vec{X_1}) \vee \ldots \vee p_n(\vec{X_n}) \leftarrow$\\ $r_1(\vec{Y_1}), \ldots, r_m(\vec{Y_m}), s_1(\vec{Z_1}), \ldots, s_k(\vec{Z_k}), not$ $u_1(\vec{W_1}), \ldots, not$ $u_h(\vec{W_h})$
\end{center}
be a rule $R$ belonging to $\mathcal{L}$.
We define a \emph{downward refinement operator} $\rho^{\neg\vee}$ for
$(\mathcal{L}, \succeq_{\mathcal{K}})$ such that the set
$\rho^{\neg\vee}(R)$ contains all $R^\prime \in\mathcal{L}$ that can be obtained from $R$ by
applying one of the following refinement rules:
\begin{description}
  \item [$\langle AddDataLit\_B^{+} \rangle$] $body(R^\prime)=body(R) \cup \{r_{m+1}(\vec{Y_{m+1}})\}$ if
		\begin{enumerate}
			\item $r_{m+1} \in P_\textsc{D}^{+}(\mathcal{L})$
			\item $r_{m+1}(\vec{Y_{m+1}})\not\in body(R)$
		\end{enumerate}
\end{description}
\begin{description}
  \item [$\langle AddDataLit\_B^{-} \rangle$] $body(R^\prime)=body(R) \cup \{$$not$ $u_{m+1}(\vec{W_{h+1}})\}$ if
		\begin{enumerate}
			\item $u_{h+1} \in P_\textsc{D}^{-}(\mathcal{L})$
			\item $u_{h+1}(\vec{W_{h+1}})\not\in body(R)$
		\end{enumerate}
\end{description}
\begin{description}
  \item [$\langle AddOntoLit\_B \rangle$] $body(R^\prime)=body(R) \cup \{s_{k+1}(\vec{Z_{k+1}})\}$ if
  	\begin{enumerate}
			\item $s_{k+1} \in P_\mathcal{C}(\mathcal{L}) \cup P_\mathcal{R}(\mathcal{L})$
			\item it does not exist any $s_{l}\in body(H)$ such that $s_{k+1} \sqsubseteq s_{l}$
		\end{enumerate}
\end{description}
\begin{description}
  \item [$\langle SpecOntoLit\_B \rangle$] $body(R^\prime)= (body(R) \setminus \{s_{l}(\vec{Z_{l}})\}) \cup s_{l}^\prime(\vec{Z_{l}})$ if
  	\begin{enumerate}
			\item $s_{l}^\prime \in P_\mathcal{C}(\mathcal{L}) \cup P_\mathcal{R}(\mathcal{L})$
			\item $s_{l}^\prime \sqsubseteq s_{l}$
		\end{enumerate}
\end{description}
\begin{description}
  \item [$\langle AddDataLit\_H \rangle$] $head(R^\prime)=head(R) \cup \{p_{n+1}(\vec{X_{n+1}})\}$ if
		\begin{enumerate}
			\item $p_{n+1} \in P_\textsc{D}^{+}(\mathcal{L})$
			\item $p_{n+1}(\vec{X_{n+1}})\not\in head(R)$
		\end{enumerate}
\end{description}
\begin{description}
  \item [$\langle AddOntoLit\_H \rangle$] $head(R^\prime)=head(R) \cup \{p_{n+1}(\vec{X_{n+1}})\}$ if
   	\begin{enumerate}
			\item $p_{n+1} \in P_\mathcal{C}(\mathcal{L}) \cup P_\mathcal{R}(\mathcal{L})$
			\item it does not exist any $p_{i}\in head(R)$ such that $p_{n+1} \sqsubseteq p_{i}$
		\end{enumerate}
\end{description}
\begin{description}
  \item[$\langle GenOntoLit\_H \rangle$] $head(R^\prime)=(head(R) \setminus \{p_{i}(\vec{X_{i}})\}) \cup p_{i}^\prime(\vec{X_{i}})$ if
		\begin{enumerate}
			\item $p_{i}^\prime \in P_\mathcal{C}(\mathcal{L}) \cup P_\mathcal{R}(\mathcal{L})$
			\item $p_{i} \sqsubseteq p_{i}^\prime$
		\end{enumerate}
\end{description}
\end{definition}

Note that, since we are working under NM-semantics, two distinct rules, namely $\langle AddDataLit\_B^{-} \rangle$ and $\langle AddDataLit\_H \rangle$, are devised for adding negated \textsc{Datalog} atoms to the body and for adding \textsc{Datalog} atoms to the head, respectively. 
It can be proved that all the rules of $\rho^{\neg\vee}$ are correct, i.e. the $R^\prime$'s obtained by applying any of the
rules of $\rho^{\neg\vee}$ to $R\in \mathcal{L}$ are such that $R \succ_{\mathcal{K}}^{\neg\vee} R^\prime$. Intuitively, it is sufficient to observe that the application of any of the rules of $\rho^{\neg\vee}$ conceived to strenghten $body(R)$ reduces the number of
models of $R$ whereas the rules aiming at weakening $head(R)$, when applied, do not augment the number of models of $R$. 

\begin{example}\label{ex:ref-op-2}
From the rule belonging to the language $\mathcal{L}$ specified in Example \ref{ex:hyp-lang-2}:
\begin{tabbing}
  MM \= MMMMMM \= MM \= \kill
\> $\leftarrow$ \texttt{enrolled(X,c1)}
\end{tabbing}
we obtain the following rules by applying $\langle AddDataLit\_B^{+} \rangle$:
\begin{tabbing}
  MM \= MMMMMM \= MM \= \kill
\> $\leftarrow$ \texttt{enrolled(X,c1)}, \texttt{boy(X)}\\
\> $\leftarrow$ \texttt{enrolled(X,c1)}, \texttt{girl(X)}\\
\> $\leftarrow$ \texttt{enrolled(X,c1)}, \texttt{enrolled(X,c2)}\\
\> $\leftarrow$ \texttt{enrolled(X,c1)}, \texttt{enrolled(X,c3)}
\end{tabbing}
the following ones by applying $\langle AddDataLit\_B^{-} \rangle$:
\begin{tabbing}
  MM \= MMMMMM \= MM \= \kill
\> $\leftarrow$ \texttt{enrolled(X,c1)}, \texttt{not boy(X)}\\
\> $\leftarrow$ \texttt{enrolled(X,c1)}, \texttt{not girl(X)}
\end{tabbing}
the following ones by applying $\langle AddOntoLit\_B \rangle$:
\begin{tabbing}
  MM \= MMMMMM \= MM \= \kill
\> $\leftarrow$ \texttt{enrolled(X,c1)}, \texttt{PERSON(X)}\\
\> $\leftarrow$ \texttt{enrolled(X,c1)}, \texttt{FEMALE(X)}\\
\> $\leftarrow$ \texttt{enrolled(X,c1)}, \texttt{MALE(X)}
\end{tabbing}
the following ones by applying $\langle AddDataLit\_H \rangle$:
\begin{tabbing}
  MM \= MMMMMM \= MM \= \kill
\> \texttt{boy(X)} $\leftarrow$ \texttt{enrolled(X,c1)}\\
\> \texttt{girl(X)} $\leftarrow$ \texttt{enrolled(X,c1)}\\
\> \texttt{enrolled(X,c2)} $\leftarrow$ \texttt{enrolled(X,c1)}\\
\> \texttt{enrolled(X,c3)} $\leftarrow$ \texttt{enrolled(X,c1)}
\end{tabbing}
and the following ones:
\begin{tabbing}
  MM \= MMMMMM \= MM \= \kill
\> \texttt{PERSON(X)} $\leftarrow$ \texttt{enrolled(X,c1)}\\
\> \texttt{FEMALE(X)} $\leftarrow$ \texttt{enrolled(X,c1)}\\
\> \texttt{MALE(X)} $\leftarrow$ \texttt{enrolled(X,c1)}
\end{tabbing}
by applying $\langle AddOntoLit\_H \rangle$.
\end{example}

\subsection{The algorithm}\label{sect:claudien-like-algo}

\begin{figure}[t]
  \centering
\begin{tabbing}
    MMM \= MMM \= MMM \= MMM \= MMM \kill
    NMDISC-$\mathcal{DL}$+\textsc{log}$^{\neg\vee}$($\mathcal{L}$, $\mathcal{K}$, $\Pi_F$)\\
    1.  $\mathcal{H}\leftarrow \emptyset$\\
    2.  $\mathcal{Q}\leftarrow \{\ \square \}$\\
    3.  \textbf{while} $\mathcal{Q} \neq \emptyset$ \textbf{do}\\
    4.  \> $\mathcal{Q} \leftarrow \mathcal{Q} \setminus \{ R \}$;\\
    5.  \> \textbf{if} NMSAT-$\mathcal{DL}$+\textsc{log}($\mathcal{K} \cup \Pi_F \cup \mathcal{H} \cup \{ R \}$)\\
    6.  \> \> \textbf{then} $\mathcal{H} \leftarrow \mathcal{H} \cup \{ R \}$\\
    7.  \> \> \textbf{else} $\mathcal{Q} \leftarrow \mathcal{Q} \cup \{ R^\prime \in \mathcal{L}| R^\prime \in \rho^{\neg\vee}(R)\}$\\
    8. \> \textbf{endif}\\
    9. \textbf{endwhile}\\
    \textbf{return} $\mathcal{H}$
\end{tabbing}
  \caption{Main procedure of NMDISC-$\mathcal{DL}$+\textsc{log}$^{\neg\vee}$}\label{fig:claudien-like-algo}
\end{figure}

The integrity theory $\mathcal{H}$ we would like to discover is a set of \textsc{$\mathcal{DL}$+log}$^{\neg\vee}$ rules. It must be induced by taking the background theory $\mathcal{K}=\Sigma \cup \Pi_R$ into account so that $\mathcal{B}=(\Sigma, \Pi \cup \mathcal{H})$ is a NM-satisfiable \textsc{$\mathcal{DL}$+log}$^{\neg\vee}$ KB. The algorithm in Figure \ref{fig:claudien-like-algo} defines the main procedure of NMDISC-$\mathcal{DL}$+\textsc{log}$^{\neg\vee}$: it starts from an empty theory $\mathcal{H}$ (1), and a queue $\mathcal{Q}$ containing only the empty clause (2). It then applies a search process (3) where each element $R$ is deleted from the queue $\mathcal{Q}$ (4), and tested for satisfaction w.r.t. the data $\Pi_F$ by taking into account the background theory $\mathcal{K}$ and the current integrity theory $\mathcal{H}$ (5). Note that the NM-satisfiability test includes also the current induced theory in order to deal with the nonmonotonicity of induction in the normal ILP setting. If the rule $R$ is satisfied by the database (6), it is added to the theory (7). If the rule is violated by the database, its refinements according to $\mathcal{L}$ are considered (8). The search process terminates when $\mathcal{Q}$ becomes empty (9). Note that the algorithm does not specify the search strategy. 
In order to get a minimal theory (i.e., without redundant clauses), a pruning step and a post-processing phase can be added to NMDISC-$\mathcal{DL}$+\textsc{log}$^{\neg\vee}$ by further calling NMSAT-$\mathcal{DL}$+\textsc{log}\footnote{Based on the following consequence of the Deduction Theorem in FOL: Given a KB $\mathcal{B}$ and a rule $R$ in \textsc{$\mathcal{DL}$+log}$^{\neg\vee}$, we have that $\mathcal{B} \models R$ iff $\mathcal{B} \wedge \neg R$ is unsatisfiable.}.  

\begin{example}\label{ex:claudien-like-algo}
With reference to Example \ref{ex:ref-op-2}, the following \textsc{$\mathcal{DL}$+log}$^{\neg\vee}$ rule:
\begin{tabbing}
  MM \= MMMMMM \= MM \= \kill
\> \texttt{PERSON(X)} $\leftarrow$ \texttt{enrolled(X,c1)}
\end{tabbing}
is the only one passing the NM-satisfiability test at step (5) of the algorithm NMDISC-$\mathcal{DL}$+\textsc{log}$^{\neg\vee}$. It is added to the integrity theory. All the other rules are further refined. When the learning process ends at step (9) because the queue of rules has become empty, the integrity theory will encompass the rules reported in Example \ref{ex:hyp-lang-2} because they are satisfied by the database.  
\end{example}

\section{Related Work}\label{sect:rel-work}

Very few ILP frameworks have been proposed so far that adopt a hybrid DL-CL representation for both hypotheses and background knowledge \cite{RouveirolV2000,Kietz03,Lisi08,LisiE08-ilp}. They are less or differently expressive than the one presented in this paper.

The framework proposed in \cite{RouveirolV2000} focuses on discriminant induction and adopts the ILP setting of learning from interpretations. Hypotheses are represented as \textsc{Carin}-$\mathcal{ALN}$ non-recursive rules with a Horn literal in the head that plays the role of target concept. The coverage relation of hypotheses against examples adapts the usual one in learning from interpretations to the case of hybrid \textsc{Carin}-$\mathcal{ALN}$ BK. The generality relation for hypotheses is defined as an extension of generalized subsumption. Procedures for testing both the coverage relation and the generality relation are based on the existential entailment algorithm of \textsc{Carin}. Following \cite{RouveirolV2000}, Kietz studies the learnability of \textsc{Carin}-$\mathcal{ALN}$, thus providing a pre-processing method which enables
ILP systems to learn \textsc{Carin}-$\mathcal{ALN}$ rules \shortcite{Kietz03}. 

In \cite{Lisi08}, the representation and reasoning means come from $\mathcal{AL}$-log. Hypotheses are represented as constrained \textsc{Datalog} clauses. Note that this framework is general, meaning that it is valid whatever the scope of induction is. The generality relation for one such hypothesis language is an adaptation of generalized subsumption to the $\mathcal{AL}$-log KR framework. It gives raise to a quasi-order and can be checked with a decidable procedure based on constrained SLD-resolution. Coverage relations for both ILP settings of learning from interpretations and learning from entailment have been defined on the basis of query answering in $\mathcal{AL}$-log. 
As opposite to \cite{RouveirolV2000}, the framework has been partially implemented in an ILP system \cite{LisiM04} that supports a variant of frequent pattern discovery where rich prior conceptual knowledge is taken into account in order to find patterns at multiple levels of description granularity. 

The framework presented in \cite{LisiE08-ilp} is the closest to the present work. Indeed it faces the problem of learning in $\mathcal{DL}$+\textsc{log}, i.e. by disregarding the NM features of $\mathcal{DL}$+\textsc{log}$^{\neg\vee}$. Yet the framework is more general than the one illustrated here because two cases of rule learning are considered, one aimed at inducing rules with one \textsc{Datalog} literal in the head and the other rules with one $\mathcal{DL}$ literal in the head. The former kind of rule will enrich the \textsc{Datalog} part of the KB, whereas the latter will extend the $\mathcal{DL}$ part.

The main procedure of NMLEARN-$\mathcal{DL}$+\textsc{log}$^{\neg}$ follows the principles of FOIL \cite{Quinlan90} but shows some peculiarities due to the nature of the underlying KR framework, e.g. the setting of learning from entailment (which is more powerful than the use of extensional background theory and coverage testing), and the ordering of generalized subsumption (instead of $\theta$-subsumption).

The main procedure of NMDISC-$\mathcal{DL}$+\textsc{log}$^{\neg\vee}$ is inspired by CLAUDIEN \cite{DeRaedtB93} as for the scope of induction and the algorithm scheme but differs from it in several points, notably the adoption of (i) relative subsumption instead of $\theta$-subsumption, (ii) stable model semantics instead of completion semantics, and (iii) learning from entailment instead of learning from interpretations, to deal properly with the chosen representation formalism for both the background theory and the language of hypotheses. 

ILP has been also applied to data engineering tasks such as the interactive restructuring of databases giving rise to the so-called Inductive Data Engineering (IDE) \cite{Flach93,Flach98,SavnikF00}. The main idea is to use induction to determine integrity constraints, such as functional and multivalued dependencies, that are valid (or almost valid) in a database and then use the constraints to decompose (restructure) the database. 

\section{Conclusions and Future Work}\label{sect:concl}

In this paper, we have investigated two ILP solutions for learning in the KR framework of \textsc{$\mathcal{DL}$+log}$^{\neg\vee}$, both valid for any $\mathcal{DL}$ for which the instantiation of the framework is decidable, but one restricted to \textsc{Datalog}$^{\neg}$ and the other for the full framework. Indeed, well-known ILP techniques for induction such as the orderings of generalized subsumption and relative subsumption have been reformulated in terms of the deductive reasoning mechanims of \textsc{$\mathcal{DL}$+log}$^{\neg\vee}$, namely by relying on the algorithm NMSAT-$\mathcal{DL}$+log devised to prove NM-satisfiability of \textsc{$\mathcal{DL}$+log}$^{\neg\vee}$ KBs. Notably, we have defined generality orders, refinement operators and coverage tests on the basis of NMSAT-$\mathcal{DL}$+log. Though the work presented in this paper is not yet supported by empirical evidence, it shows that it is feasible for ILP to go beyond \textsc{Datalog} towards \textsc{$\mathcal{DL}$+log}$^{\neg\vee}$. The potential of this extended ILP has been illustrated in two traditional database problems, i.e. the definition of views and the definition of integrity theories, for which we have sketched ad-hoc ILP algorithms, NMLEARN-$\mathcal{DL}$+\textsc{log}$^{\neg}$ and NMDISC-$\mathcal{DL}$+\textsc{log}$^{\neg\vee}$, respectively. The NM features as well as the DL component of \textsc{$\mathcal{DL}$+log}$^{\neg\vee}$ enable these algorithms to build hypotheses with expressiveness far greater than the one reachable with the predecessors FOIL and CLAUDIEN. Notably, ontologies accommodate elegantly in the solution to the database problems being considered. From the ILP viewpoint the expressive power of \textsc{$\mathcal{DL}$+log}$^{\neg\vee}$ has, of course, raised some technical difficulties. In particular, the critical point has been the DL component that has required an appropriate treatment when defining both the generality orders and the refinement operators. Also the setting of learning from entailment turned out to be the most appropriate for the induction within the \textsc{$\mathcal{DL}$+log}$^{\neg\vee}$ KR framework.

As next step towards any practice, we plan to first analyze the complexity and then produce an efficient and scalable implementation of these ILP algorithms. Adopting less expressive but tractable instantiations of \textsc{$\mathcal{DL}$+log}$^{\neg\vee}$ may turn out crucial from this point of view. E.g., DL-Lite \cite{CalvaneseGLLR07} has been proved to be good at making \textsc{$\mathcal{DL}$+log}$^{\neg\vee}$ practically useful \cite{Rosati06}. Another point is the definition of so-called optimal refinement operators to be actually employed in NMLEARN-$\mathcal{DL}$+\textsc{log}$^{\neg}$ and NMDISC-$\mathcal{DL}$+\textsc{log}$^{\neg\vee}$. Indeed, ideal refinement operators are mainly of theoretical interest, because in practice they are often very inefficient. More constructive - though possibly improper - refinement operators are usually to be preferred over ideal ones. Optimal refinement operators can be easily derived from those proposed in this paper.

Learning in \textsc{$\mathcal{DL}$+log}$^{\neg\vee}$ is also promising for Semantic Web applications for the following reasons. First, it can deal with ontologies almost as expressive as the ones that OWL allow. Indeed, as already mentioned, $\mathcal{SHIQ}$ has been the starting point for the definition of OWL and gives rise to one of the currently most expressive decidable instantiations of \textsc{$\mathcal{DL}$+log}$^{\neg\vee}$. Second, it can deal with incomplete knowledge thanks to the NM features of \textsc{$\mathcal{DL}$+log}$^{\neg\vee}$. Third, it can deal with ontologies and rules tightly integrated as devised by the W3C Rule Interchange Format (RIF) working group.\footnote{\texttt{http://www.w3.org/2005/rules/wiki/RIF\_Working\_Group}}  Indeed the activity of the RIF group concerns (i) the definition of a core language with extensions some of which (the nonmonotonic ones) will most likely be inspired by hybrid DL-CL languages like \textsc{$\mathcal{DL}$+log}$^{\neg\vee}$ and (ii) the identification of use cases many of which are suitable to our algorithms for application. 

As a final remark, we would like to point out that the shift from \textsc{Datalog} to \textsc{$\mathcal{DL}$+log}$^{\neg\vee}$ in ILP paves the way to an extension of Relational Learning (and Data Mining), named Onto-Relational Learning, which accounts for ontologies in a clear, well-founded and systematic way. Following the work reported in this paper, we can build new-generation ILP systems able to learn from relational databases integrated
with ontologies according to the principles of Onto-Relational Learning.   

\smallskip 
\noindent
{\textbf{Acknowledgements} We are grateful to Riccardo Rosati for his precious advice on \textsc{$\mathcal{DL}$+log}$^{\neg\vee}$ and Diego Calvanese for his valid support on the Boolean CQ/UCQ containment problem. Also we thank the anonymous reviewers whose commments very much helped us improving this paper.


\end{document}